\documentstyle[dina4,11pt,times]{article}

\makeatletter
\def\^{{}^}
\def\_{{}\lowind1_}
\def\scr#1{{\scriptstyle #1}}
\def\lowind#1{\@ifnextchar_{\makebox[-.#1em]{ }}{}}
\def\bbar#1#2{\bar{#2\makebox[.#1em]{ }}\makebox[-.#1em]{ }}
\def\alg#1({{\bf #1}(}
\def\grp#1({{\rm #1}(}
\def\ft#1#2{{\textstyle{{#1}\over{#2}}}}
\def\cc{\makebox[0pt][l]{$^*$}}
\def\intd#1#2{\int \d^{#1}\!#2 \,}
\def\comm#1#2{\big[#1,#2\big]}
\def\pois#1#2{\big\{#1,#2\big\}}
\def\deldel#1/#2/{\frac{\del #1}{\del #2}\,}
\def\deltadelta#1/#2/{\frac{\delta #1}{\delta #2}\,}
\def\del{\partial\lowind{08}}
\def\D{D\lowind1}
\def\Dd{{\cal D}\lowind1}
\def\nabl{\nabla\lowind1}
\def\e#1#2{e_{#1}\^{#2}}
\def\ee#1#2{\tilde e_{#1}\^{#2}}
\def\E#1#2{E_{#1}\^{#2}}
\def\P#1#2{P_{#1}\^{#2}}
\def\Pcc#1#2{P\cc_{#1}\^{#2}}
\def\Q#1#2{Q_{#1}\^{#2}}
\def\AP#1#2#3#4{W_{#1#2#3}\^{#4}}
\def\APrest#1#2#3{X\cc_{#1#2}\^{#3}}
\def\follows{\qquad\Rightarrow\qquad}
\def\equivalent{\qquad\Leftrightarrow\qquad}
\def\diag{{\rm diag}}
\def\schwach{\approx}
\def\O{{\mit \Omega}}
\def\Delt{{\mit \Delta}}
\def\o{\omega}
\def\p{p}
\def\A{A}
\def\J{J}
\def\R{R}
\def\F{F}
\def\Gf{{\cal G}}
\def\LP{\lambda}
\def\LPP{\kappa}
\def\HP{V}

\def\Wir{S}
\def\Lag{L}
\def\g{u}
\def\h{v}
\def\sig{\sigma\lowind{1}}
\def\eins{{\bf 1}}
\def\eps{\varepsilon}
\def\i{{\rm i}}
\def\d{{\rm d}}
\def\Tr{\,{\rm Tr}}
\def\grav{\psi\lowind1}
\def\bgrav{\bar\psi\lowind1}
\def\pgrav{\bar\pi}
\def\seps{\epsilon}
\def\bseps{\bar\epsilon}
\def\fmod{\phi}
\def\tildePsi{\widetilde \Psi}
\def\tildePhi{\widetilde \Phi}
\def\bm#1{\mbox{\boldmath$#1$}}
\def\L{{\bm L}}
\def\H{{\bm H}\lowind{15}}
\def\S{{\bm S}\lowind1}
\def\bS{\bbar2{\bm S}\lowind1}
\def\RR{{\sf R}}
\def\CC{{\sf C}}
\def\x{{\rm x}}
\def\y{{\rm y}}
\def\z{{\rm z}}
\def\t{{\,\rm t}}
\def\X{x}

\def\xx{x}
\def\yy{y}
\def\ket#1{\vert\,#1\,\rangle}

\def\braket#1#2{\langle\,#1\,\vert\,#2\,\rangle}
\def\Hphys{{\cal H}}
\def\Lphys{{\cal L}}
\def\Sphys{{\cal P}}
\def\gllabel[#1]{\label{#1}}
\def\beq{\arraycolsep.1em\begin{eqnarray}\@ifnextchar[{\gllabel}{}}
\def\eeq{\end{eqnarray}}
\def\zl{\nonumber\\}

\@addtoreset{equation}{section}
\def\eref#1{(\ref{#1})}
\def\sref#1{section~\ref{#1}}
\makeatother

\begin{document}
\title{New representation and a vacuum state\\for canonical quantum gravity}
\author{Hans-J\"urgen Matschull\\
        Department of Mathematics, King's College London,\\
        Strand, London WC2R 2LS, England\\
        email: hjm@mth.kcl.ac.uk}
\date{kcl-th-94-22, gr-qc/9412020\\6 December 1994}
\maketitle

\begin{abstract}
\noindent A new representation for canonical gravity and supergravity is
presented, which combines advantages of Ashtekar's and the Wheeler~DeWitt
representation: it has a nice geometric structure and the singular metric
problem is absent. A formal state functional can be given, which has some
typical features of a vacuum state in quantum field theory. It can be
canonically transformed into the metric representation. Transforming the
constraints too, one recovers the Wheeler~DeWitt equation up to an anomalous
term. A modified Dirac quantization is proposed to handle possible anomalies in
the constraint algebra.
\end{abstract}

\section{The classical action\label{class}}
The easiest way to obtain the new representation of canonical gravity is to
start from the complex Lagrangian for general relativity which can also be used
to derive Ashtekar's variables and the polynomial constraints directly from an
action principle~\cite{ashtekar:86,jacobson.smolin:88b}. The basic field
variables appearing in this action are the vierbein components $\E MA$ with
covector index $\scr M$, taking the values $\t,\x,\y,\z$ for the local
coordinates, and the flat index $\scr A=0,1,2,3$, raised and lowered using the
Minkowski metric $\eta_{AB}=\diag(-1,1,1,1)$. In addition, the ``Ashtekar
action'' depends on the  $\alg so(3,\CC)$ connection $\A_{Ma}$, where $a=1,2,3$
labels the generators of $\alg so(3)$. It can be interpreted as the $\alg
so(3,\CC)$ representation of the $\alg so(1,3)$ spin connection $\O_{MAB}$. The
algebras $\alg so(1,3)\simeq\alg so(3,\CC)$ are mapped onto each other by an
isomorphism
\beq[alg-map]
&\A_{Ma} = \J_a\^{AB} \O_{MAB}, \qquad
\A\cc_{Ma} = \J\cc_a\^{AB} \O_{MAB}& \zl
&\O_{MAB} = \A_{Ma} \J_{aAB} + \A\cc_{Ma} \J\cc_{aAB},&
\eeq
where $\J_{aAB}$ is a constant ``matrix'' mapping antisymmetric real tensors
onto complex 3-vectors. Properties of these $\J$-symbols are summarized in the
appendix.

The difference between Ashtekar's representation and that presented here is
that we will not treat the connection as an independent field and thus it will
not appear as a canonical variable or quantum operator (however, it will appear
as a useful function on phase space later on). Instead, it is defined as a
function of the vierbein and its derivatives, implicitly given by the vierbein
postulate, i.e.\ it is required that the vierbein is covariantly constant. The
equations are most simply written as
\beq[vierb-post]
\D_{[M} \E{N]}A = \del_{[M} \E{N]}A + \O_{[M}\^A\_B \E{N]}B = 0 .
\eeq
As is well known this defines $\O_{MAB}[E]$ uniquely if and only if the
vierbein is invertable (see e.g.~\cite{matschull.nicolai:92} for the explicit
solution).

We will therefore assume that the determinant $E=\det(\E MA)$ does not vanish
and the inverse vierbein $\E AM$ exists. This is another crucial difference
between this and Ashtekar's representation: there one has to give up this
restriction to define the connection representation properly, but on the other
hand one must use it first to obtain the polynomial constraints
(see~\cite{matschull:94a} for a critical discussion of these problems). Here we
will insist on invertable metrics, which also has the consequence that there is
no need to have polynomial constraints: phase space functions like $E^{-1}$ are
well defined.

The Einstein Hilbert action can be expressed in terms of the field strength of
the spin connection
\beq[action]
\Wir[E] =  - \ft \i2 \intd4\X \eps^{MNPQ}\, \E MA \E NB \J_{aAB} \F_{PQa}[E],
\eeq
where
\beq
  \F_{PQa} = \del_P \A_{Qa} - \del_Q \A_{Pa} + \eps_{abc} \A_{Pb} \A_{Qc}.
\eeq
An explicit derivation is given in the appendix.

Remember that this so called 1.5 order action has the useful property that the
spin connection obeys its own equation of motion, which is just the vierbein
postulate, and this remains true when writing the action in terms of $\A_{Ma}$
instead of $\O_{MAB}$ (this is not as trivial as one might think; see the
appendix for a proof). Whenever we compute functional derivatives of the action
with respect to $\E MA$, we only have to vary the explicit vierbein fields
appearing in~\eref{action}, as long as we do not use the vierbein postulate
before calculating the derivative.

The basic gauge symmetries of the Einstein Hilbert action are, of course, the
local Lorentz symmetry acting on the flat indices, and the invariance under
diffeomorphisms of the background 4-manifold. Under a local Lorentz
transformation with parameter $\LP_a=\J_a\^{AB}\LP_{AB}$ the fields transform
as
\beq[lorentz]
\delta \E MA = \LP^A\_B \E NB \follows
  \delta \O_{MAB} &=& - \D_M \LP_{AB}, \zl
  \delta \A_{Ma} &=& - \D_M \LP_a.
\eeq
Note that the vierbein is the only primary field here and thus the
transformations of $\O_{MAB}$ and $\A_{Ma}$ are obtained via~\eref{vierb-post}.
Of course, all this is well known, but let us explain the main idea of this
article. In Ashtekar's representation the basic phase space variable is
$\A_{Ma}$, which is the $\alg so(3,\CC)$ gauge field of this symmetry. The
constraint associated with this gauge freedom is easily solved by considering
Wilson loops~\cite{jacobson.smolin:88}. There has been much effort to construct
new kinds of representations based on these invariants, and to solve the
remaining contraints. However, many problems of this ``Ashtekar programm'',
which were known from the very beginning, are still unsolved, e.g.\ how to
treat singular metrics~\cite{matschull:94a}.

Ashtekar's representation splits the contraints, and thus the gauge symmetries,
into a simple part, the Lorentz transformations, and a more complicated part
consisting in principle of the generators of the four dimensional
diffeomorphisms. These are hard to solve, mainly because the diffeomorphism
group is somewhat awkward to deal with. The question arising is whether it is
possible to interchange the roles of the two local gauge groups. It would then
be natural to choose a representation where $\E MA$ is the basic configuration
variable (i.e.\ the wave functional depends on $\E MA$), as in a certain sence
the metric or the vierbein may be considered as the gauge field of the
diffeomorphism group. This would also avoid the difficulties concerning the
singular metrics, because the configuration space, which is the support of the
wave functional, could be taken to be the set of all invertable vierbein
fields.

To exploit this idea, let us check whether we can identify a gauge field of the
diffeomorphism group explicitly. It must be a one form with an extra four
dimensional index, because the generator is a local 4-vector. Obviously, $\E
MA$ is such a one-form. If it is a gauge field, then there should be a symmetry
under which it transforms as
\beq[trans]
\delta \E MA = - \D_M \HP^A ,
\eeq
where $\HP^A$ is the parameter field. In fact, we found a symmetry of the
action, as can be seen easily by using the 1.5 order trick. We only have to
vary the two vierbein fields in~\eref{action}, integrate by parts, using the
vierbein postulate~\eref{vierb-post} and the Bianchi identity for the field
strength
\beq
\D_{[M} \F_{NP]a} = 0,
\eeq
which holds independently of the vierbein postulate.
Let us call~\eref{trans} a ``translation'', because it is related to the
Lorentz rotation~\eref{lorentz} like the translations of the Poincar\'e group
are related to the rotations. Explicitly, the commutator of a translation
$\HP^A$ and a rotation $\LP_{AB}$ gives another translation with parameter
$\HP^A\LP_A\^B$. However, two translations do not commute, because the spin
connection appears in~\eref{trans}.

To get the commutator of two translations, we have to compute the
transformation of $\O_{MAB}$ under a translation. Using the vierbein postulate
and some properties of the Riemann tensor, we find
\beq
\delta \O_{MAB} = \R_{MNAB} \HP^N,
\eeq
where $\HP^N=\E AN \HP^A$. Acting on~\eref{trans} with another translation
yields the commutator
\beq[trans-comm]
\comm{\delta_1}{\delta_2}\, \E MA =
\R_{PQAB} \, \HP_1^P \HP_2^Q \, \E MB,
\eeq
which is a Lorentz transfomation with field dependent parameter
$\LP_{AB}=\R_{PQAB} \HP_1^P \HP_2^Q$ or $\LP_a =\F_{PQa} \HP_1^P \HP_2^Q$.

Finally, let us see how the translations are related to the usual Lie
derivative appearing as the generator of diffeomorphisms, which has as its
parameter a tangent vector $\HP^M$. To obtain the Lie derivative, one has to
add a translation with parameter $\HP^A=\E MA\HP^M$ and a Lorentz rotation with
parameter $\LP_{AB}=\O_{MAB}\HP^M$. Using the explicit solution $\O_M\^A\_B=\E
NA \nabl_M \E BN$ for the vierbein postulate (where $\nabl_M$ is the metric
covariant derivative), the transformation becomes
\beq
\delta \E MA   &=& - \D_M \big( \E NA \HP^N \big)
               + \O_N\^A\_B \HP^N \E MB \zl
  &=& - \E NA \nabl_M \HP^N - \HP^N \nabl_N \E MA.
\eeq
This is the Lie derivative of $\E MA$ along $-\HP^N$. As the vierbein is
invertable, there is a one-to-one relation between translations and generators
of diffeomorphisms and we may regard the translations as the basic symmetries
of the action instead of the diffeomorphisms.

\section{Canonical formulation\label{canon}}
Here we will derive the classical constraint algebra by applying the Dirac
canonical formalism~\cite{dirac:65} to the action~\eref{action}. If we use the
vierbein components as canonical configuration variables and Langrange
multipliers, instead of introducing a lapse function and a shift vector, we end
up with a Lorentz covariant set of hamiltonian constraints. They have a nice
geometric structure like Ashtekar's constraints, but do not split into a vector
and scalar constraint.

\subsection*{Space time split, Lagrangian and momenta}
We will now  Space time is split into a three dimensional space spanned by
coordinates $m=\x,\y,\z$ and time $\t$, which must be assumed to be a global
coordinate. Thus all space time indices split into $\scr M \mapsto m,\t$. If we
define the spacial Levi Civita tensor by $\eps^{mnp}=\eps^{\t mnp}$ (or
$\eps^{\x\y\z}=\eps_{\x\y\z}=1$), then the Lagrange density becomes
\beq[lag-1]
       - \i \eps^{mnp} \, \E mA \E nB  \J_{aAB} \F_{\t pa}[E]
       - \i \eps^{mnp} \, \E \t A \E mB  \J_{aAB} \F_{npa}[E].
\eeq
Note that we do not fix any gauge here, i.e.\ the vierbein is not required to
split into a dreibein, a lapse function and a shift vector. The Lagrangian is
still invariant under the full local Lorentz group. However, let us impose a
restriction on the configuration space: we require the hypersurface to be
spacelike, i.e.\ its intrinsic metric $g_{mn}=\E mA\E nB\eta_{AB}$ must have
signature $(+,+,+)$. It is important to note that this restiction has nothing
to do with a gauge fixing of the local Lorentz group. It is the usual
restriction one imposes on the metric in the
Wheeler~DeWitt~\cite{wheeler:64,dewitt:67} approach, but replacing the metric
by the vierbein does not automatically lead to a vierbein splitting into a
dreibein, a lapse and a shift vector.

Because of the fixed signature of $\eta_{AB}$ it is equivalent to require
$g=\det(g_{mn})>0$, or that there exists a timelike normal vector $N^A$
uniquely defined by
\beq[N-def]
\eps_{ABCD} N^A \E mB \E nC \E pD = \eps_{mnp} , \qquad
N_A \E mA = 0.
\eeq
Observe that $N^A$ is a density of weight $-1$ under diffeomorphisms and that
is does not exist if we allow the hypersurface to become lightlike, as the
normal vector then becomes tangent to the surface itself and cannot be
normalized by the first equation in~\eref{N-def}. The restriction of the
configuration space will simplify the discussions below, where we will always
assume that $N^A$ exists.

The expression~\eref{lag-1} for the Lagrange density still contains the second
time derivative of the vierbein in its first term, which must be eliminated by
a partial integration. Thus we define
\beq[lag-2]
    \Lag = \intd3\xx \Big(
  2\i\eps^{mnp}\, \del_\t\E mA\,\E nB\J_{aAB}\A_{pa} -\ \ \ \ \ && \zl
{}- \i \eps^{mnp} \, \D_p \big( \E mA \E nB \J_{aAB} \big) \,\A_{\t a}
 &-& \i \eps^{mnp} \, \E \t A \E mB \J_{aAB} \F_{npa}   \Big).
\eeq
{}From now on the action is different from the Einstein Hilbert action, because
we added a complex total derivative, thus the action itself is complex and one
has to deal with complex momenta obeying certain reality conditions. How to do
this if the imaginary part of the action is a total divergence has been worked
out in~\cite{fukuyama.kamimura:90}. Let us first obtain the momenta and then
derive the primary constraints and reality conditions they obey.

The connection components $\A_{ma}$ and $\A_{\t a}$ still obey their own
equations of motion, thus we can neglect their dependence on $\del_\t \E mA$
when differentiating the action: the momentum conjugate to $\E mA$ reads
\beq[P-def]
\P Am[E,\del_\t E]
 = 2 \i \eps^{mnp} \, \J_{aAB} \, \E nB \A_{pa}[E,\del_\t E]
\eeq
and that conjugate to $\E \t A$ vanishes, i.e.\ the time components of the
gauge fields are Lagrange multipiers as they should be. The phase space is
given by the tangent bundle of the configuration space defined above, and the
Poisson brackets read
\beq[pois]
\pois{ \E mA }{ \P Bn } = \delta_m^n \, \delta_B^A.
\eeq
Whenever such a bracket or quantum commutator appears, the dependence of the
fields on the space points and the spacial delta function will not be written
out, as long as no derivatives are involved and it is obvious how to restore
them: $\{A,B\}=C$ has to be read as $\{A(\xx),B(\yy)\}=C(\xx)\delta(\xx,\yy)$
if $A,B,C$ are local fields.

Before considering the constraints let us discuss the reality conditions on
these momenta. They are obviously complex but are conjugate to real variables;
there should be a relation giving $\Pcc Am$ as a holomorphic function of $\P
Am$ and the spacial components of the vierbein. Calculating the imaginary part
of the momentum explicitly, we find that
\beq
\P Am - \Pcc Am   &=&
2\i \eps^{mnp} \, \E nB \big( \J_{aAB} \A_{pa} + \J\cc_{aAB} \A\cc_{pa}
\big)\zl
  &=& 2 \i \eps^{mnp} \, \E nB \O_{pAB}
 = 2 \i \eps^{mnp} \, \del_n E_{pA}.
\eeq
In contrast to the connection representation, where the reality constraints on
Ashtekar's variables are non-polynomial, we have a very simple linear relation
\beq[reality]
\Pcc Am = \P Am - 2 \i \eps^{mnp} \, \del_n E_{pA}.
\eeq
The relation can also be written as
\beq[Q-def]
\Q Am = \P Am -  \i \eps^{mnp} \, \del_n E_{pA} \in \RR,
\eeq
and this $\Q Am$ is nothing but the momentum of $\E mA$ that would come out if
we used the real Einstein Hilbert action instead of~\eref{lag-2}.

One can easily see that another approach to the complex momentum is to perform
a canonical transformation from $\Q Am$ to $\P Am$ using the phase space
functional (see~\cite{henneaux.schomblond.nelson:89} for the similar
construction of Ashtekar's variables)
\beq
\ft\i2 \intd3\xx  \eps^{mnp} \, \E mA \del_n E_{pA},
\eeq
whose time derivative is the imaginary part of the difference
between~\eref{lag-1} and~\eref{lag-2}. In principle we have to regard the
reality conditions as second class constraints (with conjugate constraints
$E\cc_m\^A=E_m\^A$) and compute the resulting Dirac brackets. However, one can
show that the Dirac brackets are equal to the Poisson brackets defined
by~\eref{pois} if every phase space function is expressed as a holomorphic
function of $\P Am$~\cite{fukuyama.kamimura:90,matschull:doc}.

\subsection*{The constraints}
The reality constraints are not the only relations between the momenta
following from~\eref{P-def}. There are also primary constraints. Using the fact
that $\J\cc_a$ commutes with $\J_b$, we find that
\beq[Lcca-def]
  \L\cc_a   &=& \J\cc_a\^{AB} \, \P Am \, E_{mB} \zl
            &=& 2\i \eps^{mnp} \, \J\cc_a\^A\_B \J_{bAC} \, \E mB \E nC \A_{pb}
          \schwach 0,
\eeq
because the product of the two $\J$ symbols is symmetric in $\scr B,\scr C$
(see~\eref{J-comm}). Note that these are in fact 3 complex (or 6 real)
equations {\em in addition} to the reality conditions on $\P Am$. If we compute
the complex conjugate by using~\eref{reality}, we obtain 3 new equations which
cannot be written as holomorphic functions of~\eref{Lcca-def}:
\beq[La-def]
\L_a
  &=& \J_a\^{AB} \, \big( \P Am - 2 \i \eps^{mnp} \, \del_n E_{pA} \big) E_{mB}
\zl
  &=& \J_a\^{AB} \, \P Am \, E_{mB} -
           \i \eps^{mnp} \, \del_m \big( \J_{aAB} \, \E nA \E pB \big)
\schwach 0.
\eeq
Of course, $\L_a$ and $\L\cc_a$ are the generators of Lorentz rotations, and
they can also be given in the $\alg so(1,3)$ representation
\beq[LAB-def]
\L_{AB} = \P {[A}m \, E_{mB]}  - \i \eps^{mnp} \, \J_{aAB} \,
 \del_m \big(  \J_{aCD} \, \E nC \E pD \big).
\eeq
Using the notation
\beq[L-schmier]
  \L[\LP_a] = \intd3\xx \LP_a \L_a, \qquad
 \L[\LP\cc_a] = \intd3\xx \LP\cc_a \L\cc_a, 
\eeq
we find that
\beq[L-action]
   \pois{\E mA}{\L[\LP_a]} &=& \LP_a \J_a\^A\_B \, \E mB, \zl
   \pois{\P Am}{\L[\LP_a]} &=& \LP_a \J_{aA}\^B \, \P Bm
             - 2\i\eps^{mnp} \J_{aAB} \, \E nB \, \del_p \LP_a ,  \zl
   \pois{\E mA}{\L[\LP\cc_a]} &=& \LP\cc_a \J\cc_a\^A\_B \, \E mB, \zl
   \pois{\P Am}{\L[\LP\cc_a]} &=& \LP\cc_a \J\cc_{aA}\^B \, \P Bm.
\eeq
We see already here that what we obtain is somehow a mixture of the
Wheeler~DeWitt and Ashtekar's representation. The conjugate momentum of the
vierbein transforms as a tensor under half of the Lorentz algebra, the
corresponding constraint having the form ``$E\times P$'', but as a connection
under the other half, represented by a constraint of the form ``$\del E +
E\times P$''. The brackets of $\L$ with itself form a local $\alg so(3,\CC)$
algebra:
\beq
   \pois{ \L[\LP_a] }{ \L[\LPP_a] }  &=& \L[\eps_{abc} \LP_b \LPP_c] , \zl
   \pois{ \L[\LP\cc_a] }{ \L[\LPP\cc_a]}&=&\L[\eps_{abc}\LP\cc_b\LPP\cc_c] ,
\zl
   \pois{ \L[\LP_a] }{ \L[\LPP\cc_a] }  &=& 0 .
\eeq

We should now check whether we have found all primary constraints. To see that
there are no more primary constraints, we have to show that for every pair
$E,P$ satisfying the constraints there is a velocity $\del_\t E$ such that
$P[E,\del_\t E]=P$, where $P[E,\del_\t E]$ is defined by~\eref{P-def}.

To show this, we invert the relation~\eref{P-def} to obtain a phase space
function $\A[E,P]$. A little algebra and making use of~\eref{N-def} and the
formulas for $\J$ in the appendix shows that the inverse of~\eref{P-def} is
\beq[A-von-P]
\A_{pa}[E,P] = \J_{aCD} \, \big( 2 \E pB \E qC - \E pC \E qB \big) \,
            N^D \, \P Bq.
\eeq
By inserting this into~\eref{P-def} we get $\P Am$ back, if and only if
$\L_a=0$ and $\P Am$ obeys the reality conditions. To obtain the velocities as
phase space functions, we choose arbitrary values for $\A_{\t a}$ and $\E \t A$
(such that $\E MA$ is non-singular) and define
\beq
\del_\t \E mA = \del_m \E \t A + \O_m\^A\_B \E \t B -
                                 \O_\t\^A\_B \E m B,
\eeq
where $\O$ is given by~\eref{alg-map}. As these are part of the definition
equations of $\O_{MAB}$ (or $\A_{Ma}$) as functions of $\E mA$ and its
derivatives, we just have to check that the rest of these equations are
satisfied, too. They read
\beq
\del_{[m} \E {n]}A + \O_{[m}\^A\_B \E {n]}B = 0.
\eeq
Inserting~\eref{A-von-P} here and making use of the reality
constraints~\eref{reality} shows that these equations are indeed satisfied and
that we have found all primary constraints, as the inverse of~\eref{P-def}
exists if and only if $\L_a=0$. Note, however, that we made use of the
existence of $N^A$ defined by~\eref{N-def}. If we allow the hypersurface to
become lightlike, additional primary constraints may appear because the
relation~\eref{P-def} can no longer be inverted to give~\eref{A-von-P}.

{}From now on we will regard $\A_{pa}$ as a phase space function given
by~\eref{A-von-P} for $\L_a=0$. Outside this ``constraint surface'' we are free
to define $\A_{pa}[E,P]$ arbitrarily. Each choice will lead to a different
expression for the hamiltonian constraints~\eref{H-def} below, but they will
all be equal up to something proportional to $\L_a$ or $\L\cc_a$, thus the
total set of constraints is invariant. However, let us restrict $\A_{pa}$ to be
linear in  $\P Am$ on the whole phase space, as otherwise the constraint
algebra would become unnecessarily awkward. Thus
\beq[A-von-P-2]
 \A_{pa}[E,P] = \AP pamA  \P Am,
\eeq
where the ``matrix'' is any function of $\E mA$ satisfying
\beq[AP-def]
   2\i \, \AP qbmA\eps^{mnp} \, \J_{aAB} \E nB  = \delta_q^p \,\eta_{ab},
\eeq
i.e.\ it is the ``left inverse'' of the matrix appearing in~\eref{P-def}. We
can express $\P Am$ as a function of $\A_{pa}$, which is given by~\eref{P-def}
on the constraint surface. Since it is linear, however, we know that we can
have an additional term proportional to $\L\cc_a$ only, i.e.\ we have the
following relation:
\beq
\P Am =  2 \i \eps^{mnp} \, \J_{aAB} \, \E nB \A_{pa}[E,P] +
          \APrest aAm \L\cc_a ,
\eeq
where $\APrest aAm$ depends on the choice of $\AP pamA$. Taking the bracket of
this equation with $\E qC$ we find the ``right inverse'' of the matrix
in~\eref{P-def}, which becomes a useful formula below:
\beq[APrest]
 2\i \eps^{mnp} \J_{aAB} \, \E nB \, \AP paqC = \delta_q^m \,\delta_A^C
    + \APrest aAm \J\cc_{aB}\^C \E qB.
\eeq
A possible choice for $\A_{pa}[E,P]$ is~\eref{A-von-P}, but we may add any term
proportional to $\L\cc_a$. We will also assume that $\AP pamA$ transforms
properly under the Lorentz algebra as its indices indicate, and that it is
local. By using~\eref{reality} and~\eref{alg-map} we obtain $\A\cc_{pa}[E,P]$
and $\O_{mAB}[E,P]$, and covariant derivatives like~\eref{cov-vec}
and~\eref{cov-tens} are defined as phase space functions. Under these covariant
derivatives the tensors $\eta_{AB}$, $\eps_{ABCD}$, $\J_{aAB}$ and
$\J\cc_{aAB}$ are still constant, but the vierbein postulate
\beq[phase-post]
    \eps^{mnp} \D_m \E nA \schwach 0 ,
\eeq
only holds up to terms proportional to $\L_{AB}$. For the special
choice~\eref{A-von-P}, e.g., we find that
\beq
 \eps^{mnp} \D_n \E pA = \eps^{mnp} \, N^A \E nB \E pC \, \L_{BC}.
\eeq
{}From~\eref{AP-def} and~\eref{L-action} we infer that $\A_{pa}$ transforms as
a connection under self-dual Lorentz transformations:
\beq
\pois{\A_{pa}}{\L[\LP_a]} =  - \D_p \LP_a , \qquad
\pois{\A_{pa}}{\L[\LP\cc_a]} =  0.
\eeq
The remaining secondary constraints are now obtained by differentiating $\Lag$
with respect to $\E \t A$. They read
\beq[H-def]
\H_A = - \i \eps^{mnp} \, \J_{aAB} \, \E mB \, \F_{npa}.
\eeq
We will call them ``hamiltonian constraints''. They are obviously related to
the usual Wheeler~DeWitt hamiltonian and diffeomorphism constraints, and we
will show in \sref{transit} how those may be obtained from $\H_A$. Observe that
our constraints have, like Ashtekar's, a very simple geometrical structure. In
particular, they are, as opposed to the hamiltonian constraint in the metric
representation, in some sense ``more homogeneous''. They are not given as the
sum of a momentum term and a curvature term. As in Ashtekar's representation,
the two parts are combined into a term containing the 4-curvature, whereas in
the Wheeler~DeWitt equation~\eref{wdw} the intrinsic 3-curvature appears. They
are slightly more complicated than Ashtekar's hamiltonian constraint, as
$\A_{pa}$ is not a primary phase space coordinate but given
by~\eref{A-von-P-2}.

On the other hand we have much simpler reality conditions on the variables,
which, in addition, asign non-trivial (but linear) conjugacy relations to the
{\em momentum}, which will appear as a derivative operator in quantum theory.
In Ashtekar's representation it is the multiplication operator whose complex
conjugate is a non-polynomial function of the derivative operator, and due to
this fact it might be much harder to solve the problem of the scalar product on
the state space than it is in the metric representation.

Our hamiltonian constraints are complex too, but they represent four real
constraints only. Computing the imaginary part gives
\beq
\H_A - \H\cc_A = -\i \eps^{mnp} \, \E mB \, \R_{npAB} =
  - 2\i \eps^{mnp} \D_n \D_p E_{mA} \schwach 0,
\eeq
i.e.\ it is proportional to the Lorentz constraints and $\H\cc_A$ is a
holomorphic linear function of $\H_A$ and $\L_{AB}$. For quantum theory this
means that we have to solve $\L_a$ and $\L\cc_a$, but only $\H_A$ ({\em or}
$\H\cc_A$). Again, we define the smeared version
\beq[H-schmier]
\H[\HP^A] = \intd3\xx \HP^A \H_A.
\eeq
To see that this constraint generates the translations we found as symmetries
of the action, let us compute the Poisson bracket with the vierbein:
\beq
  \pois {\E qC}{\H[\HP^A]}   &=&
   2 \i \eps^{mnp} \, \D_n \big( \J_{aAB} \HP^A \E mB \big) \,
         \AP paqC \zl
     &\schwach& - \D_q \HP^C
   - \D_m \HP^A \, \APrest aAm \J\cc_{aB}\^C \E qB,
\eeq
where we used~\eref{APrest} and~\eref{phase-post}, i.e.\ we neglected terms
proportional to $\L\cc_a$. So in fact $\H_A$ generates the
translations~\eref{trans}, but in addition it generates an antiself-dual
Lorentz rotation. The brackets of $\H_A$ with the Lorentz constraints are
easily calculated, as $\H_A$ transforms covariantly under Lorentz
transformations:
\beq
  \pois{ \L[\LP_{AB}] }{ \H[\HP^A] } = \H[\LP^A\_B \HP^B ] .
\eeq
To calculate the bracket of $\H_A$ with itself is rather cumbersome. What we
would expect is to get the Lorentz constraint, as we saw in~\eref{trans-comm}
that the commutator of two translations gave a Lorentz rotation. However, as
$\H_A$ generates extra Lorentz rotations, this might not be the case here.
Instead of computing the commutator explicitly, in \sref{transit} we will use
results known from other representations to show that the classical algebra
closes and, in addition, that also the quantum algebra formally closes, if we
choose a special ordering.

\section{Quantum theory\label{quant}}
We shall now define a quantum representation and construct a formal solution to
the resulting constraints, which has some typical features of a vacuum
functional in quantum field theory. The primary field operators are the
vierbein $\E mA$ and its momentum $\P Am$. As we had to assume that $\E mA$ is
non-singular to obtain the contraint algebra, we should choose the
$E$-representation here, i.e.\ the wave functional $\Psi$ will be a function on
the set of all vierbein fields $\E mA$ with positive definit spacial metric
$g_{mn}$, and the momentum operator represented by a functional derivative.
Every other representation produces difficulties with the implementation of
these restrictions, as, e.g., the Ashtekar representation, where $\A_{pa}$
becomes a multiplication operator~\cite{matschull:94a}. In the representation
chosen here no such problems occur, the vierbein operator is still non-singular
in the sense that there exists a well defined operator for $N^A$
obeying~\eref{N-def}.

Other typical problems are, of course, still present, like ill-defined operator
products etc. We will not discuss any special regularization here, but we will
see in the end that the formal solution found suggests that standard
regularization methods could provide a well defined constraint algebra and a
well defined state functional. In \sref{transit} we will also see that other
quantization methods may be able to deal with anomalies arising from formally
ill-defined products without any regularization.

\subsection*{The constraints}
We define the operators such that their commutator is $-\i$ times the classical
Poisson bracket, thus we have
\beq[op-def]
\P Am(\xx) = \i \deltadelta / \E mA(\xx) / , \qquad
\A_{pa}(\xx) = \i \AP pamA(\xx) \deltadelta / \E mA(\xx) /,
\eeq
both acting on functionals $\Psi[\E mA]$. For the contraints we have to choose
an operator ordering. The most obvious would be to order all functional
derivatives to the right, as this is the ``less singular'' one, in the sense
that as few ill-defined operator products as possible appear. However, this
would distroy the nice geometrical structure of the hamiltonian constraint, as
it could no longer be expressed in terms of the field strength of some
connection.

Of course, another obvious choice is to take the constraints as they are given
in~\eref{H-def} and just insert the operators~\eref{op-def}. The only remaining
freedom is then where to put the vierbein in $\H_A$. Let us place it to the
left, as we will see that this leads to a formally closed algebra. There are no
ordering ambiguities in $\L_a$ or $\L\cc_a$, so the complete set of constraints
is
\beq[q-const]
\L\cc_a   &=& \J\cc_a\^{AB} \, \P Am \, E_{mB}, \zl
\L_a      &=& \J_a\^{AB} \, \P Am \, E_{mB} -
           \i \eps^{mnp} \, \del_m \big( \J_{aAB} \, \E nA \E pB \big) , \zl
\H_A      &=& -\i\eps^{mnp} \J_{aAB} \, \E mB \, \F_{npa}.
\eeq
Note that written as operators the classically complex conjugate constraints
$\L_a$ and $\L\cc_a$ are completely independent. There is no $*$-relation
between operators until a scalar product is defined. Even then the relation
exist between observables only as a scalar product such that the classical
$*$-relations are preserved can be defined on the physical phase space only.
The complex conjugate of $\H_A$, however, is still a linear combination of
$\H_A$ and $\L_{AB}$ as the reality condition on $\P Am$ holds as a second
class constraint and therefore as an exact operator identity. It has nothing to
do with an adjointness relation with respect to any scalar product on state
space.

\subsection*{The vacuum state}
We will now show that there exists a formal solution to all constraints. We
will solve the constraints step by step, but not in the order one usually does
it in the Ashtekar representation, where one first solves $\L_a$ in general and
then tries to solve $\H_A$. As already mentioned in the beginning, we want to
interchange the roles of Lorentz and diffeomorphism generators, thus we will
solve $\H_A$ first and then $\L_a$. Nevertheless let us start with $\L\cc_a$.

The complete solution to $\L\cc_a \Psi=0$ can be found easily, as this just
requires $\Psi$ to be invariant under antiself-dual Lorentz transformations.
Speaking somewhat sloppy, to provide a function of $\E mA$ that is invariant
under these transformations, we have to contract the $\scr A$ indices
completely and such that no ${}\cc_a$-index appears. The only tensors we have
to achieve this are $\eta_{AB}$, $\J_{aAB}$ and $\eps_{ABCD}$, and any
contraction of two of them is again a linear combination. So every function
invariant under antiself-dual Lorentz transformation can be expressed as a
(holomorphic) function of
\beq[ee-def]
\ee ap = - \eps^{mnp} \J_{aAB} \, \E mA \E nB \quad {\rm and} \quad
g_{mn} = \eta_{AB} \, \E mA \E nB.
\eeq
It can, in fact, be expressed as a function of $\ee am$ alone, as
\beq[g-von-ee]
\ee am \ee an = \ft12 \eps^{mpq} \eps^{nrs} g_{pr} g_{qs} = g g^{mn}
\eeq
is the densitized inverse of the three metric, thus $g_{mn}$ is determined by
$\ee am$ up to sign, which, however, is fixed by $g=\det(g_{mn})>0$. We can
solve the antiself-dual Lorentz constraint by
\beq
  \Psi = \Psi[\ee ap],
\eeq
where $\Psi$ is an arbitrary holomorphic functional.

Next we consider the equation $\H_A\Psi=0$. We need to know how $\A_{pa}$ acts
on a functional of $\ee ap$:
\beq
 \A_{pa} \, \Psi   &=& \i \AP pamA \, \deldel \ee bq / \E mA /
                        \deltadelta \Psi / \ee bq /  \zl
   &=& -2\i  \eps^{mnq} \J_{bAB} \, \E nB \, \AP pamA \,
                         \deltadelta \Psi / \ee bq /
  = -\deltadelta \Psi / \ee ap /  ,
\eeq
where we used~\eref{AP-def} and did not write out the dependence on the point
$\xx$ explicitly as all the fields are to be taken at the same point. Thus we
recover the ``dual'' of Ashtekar's representation, where $\ee am$ and $\A_{pa}$
are canonically conjugate quantities, but only after solving $\L\cc_a$.

Having this simple representation for $\A_{pa}$ we can now seek for solutions
to $\H_A\Psi=0$. As we certainly cannot find the general solution, let us look
for simple solutions. A subset of all solutions is given by the wave
functionals annihilated by $\F_{npa}$, also containing the trivial solution
$\Psi=1$. This subset can indeed be given completely. If the connection
$\A_{pa}$ is curvature free, then it is (locally, but let us assume trivial
topology of space time here) given by
\beq[A-curv-free]
  -\ft\i2 \A_{pa} \, \sig_a  = \g^{-1} \del_p \g,
\eeq
where $-\ft\i2\sig_a$ is any matrix representation of $\alg so(3,\CC)$ and $\g$
is a  matrix field taking values in the corresponding group representation. Let
us choose the $\alg su(2,\CC)$ representation here, thus $\sig_a$ are the pauli
matrices and $\g\in \grp SU(2,\CC)$. For a given field $\g$ one can define a
wave functional $\Psi_\g$ solving the corresponding quantum eigenvalue equation
\beq[A-eigenwert]
  \A_{pa} \, \Psi_\g = \i \Tr(\g^{-1} \del_p \g \sig_a) \, \Psi_\g,
\eeq
which is explicitly given by
\beq[Psi-g-def]
  \Psi_\g = \exp \Big\{ -\i \intd3\xx \Tr\big(\g^{-1}\del_p \g \sig_a \big)
                                   \, \ee ap \Big\}.
\eeq
For any field $\g$ we now have a solution to the hamiltonian constraint
$\H_A\Psi_\g=0$, which is well defined as long as $\g$ satisfies some fall-off
conditions at spacial infinity. It does not reqiure any regularization for the
constraint and is an exact solution to $\H_A$. It becomes formal, however, if
we now try to solve the self-dual Lorentz constraint.

Because of the inhomogeneous term $\L_a$ does not require $\Psi$ to be
invariant under self-dual Lorentz transformations, in contrast to the same
operator in Ashtekar's representation, where it acts as a linear differential
operator. Let us see how $\L_a$ acts on $\Psi_\g$. A short calculation
yields~\cite{matschull:doc}
\beq
\L[\LP_a] \Psi_\g = \ft\i2 \intd3\xx
         \LP_a \, \Tr\Big( \g \sig_a \deltadelta \Psi_\g / \g /\Big) ,
\eeq
where the matrix valued derivative $\del/\del\g$ is defined by
$(\del/\del\g)_{\alpha\beta}=\del/\del(\g_{\beta\alpha})$, and is equal to the
(formal) derivative if it acts on a function given by a power series such as
$\Psi_\g$, i.e.\ $\del/\del\g \Tr(X\g)=X$. Therefore $\L_a$ acting on $\Psi_\g$
generates multiplication with $\sig_a$ from the right on $\g$. It is useful to
exponentiate this relation, which gives
\beq[L-endl]
\exp\big(\L[\LP_a]\big) \, \Psi_\g = \Psi_{\g\h}, \quad {\rm where} \quad
\h = \exp\big( \ft\i2  \LP_a \, \sig_a \big).
\eeq
A formal solution to $\L[\LP_a]\Psi=0$ can now be given by integrating
$\Psi_{\g\h}$ over $\h$:
\beq[solution]
\Phi_\g = \int [\d\h] \, \Psi_{\g\h}.
\eeq
Assuming that the measure $[\d\h]$ is invariant under multiplication from the
right, we obviously have found a solution to all the constraints. As such a
measure, of course, does not exist on the space of all fields $\h$, the
solution becomes formal.

To check ``how formal'' it is, i.e.\ what kind of regularization is able to
provide a well-defined functional, it is important to note that it is
sufficient to solve $\L[\LP_a]\Psi=0$ for real $\LP_a$ only, or for $\LP_a$
element of any real subspace of $\CC^3$, as $\L_a\Psi=0$ implies
$\i\L_a\Psi=0$. In other words, it is sufficient to integrate over any real
form of $\grp SU(2,\CC)$ in~\eref{solution}. To get a well-defined integral the
best choice is, of course, to integrate over the compact real form $\grp
SU(2)$. If we then, in addition, assume that space is compact and regularize
the theory on a lattice with finitely many points, the wave function becomes
well defined, as it is given by finitely many integrations over the compact
group $\grp SU(2)$. This argument is slightly heuristic, of course, as we have
to transfer all the expressions above onto the lattice first and then check
whether a solutions like~\eref{solution} exists. Note, however, that the
problem here is much simpler than that arising in

 the loop representation, where on has to integrate over the set of all loops
or the diffeomorphism group of the three dimensional space, which are much
harder to deal with than the local $\grp SU(2)$ here. In the next section we
will see that other quantization methods may lead to a perfectly well defined
``state'', which does not require any functional integration.

Let us discuss, also a little bit heuristically, the properties of the
functional $\Phi_\g$. The first question is: how many solutions did we find?
Obviously we do not necessarily get distinct solutions for different fields
$\g$, as $\Phi_\g=\Phi_{\g\h}$ for any $\grp SU(2)$-valued field $\h$. For
$\Phi_\g \ne \Phi_{\g'}$ we must have $\g^{-1}\g'\not\in\grp SU(2)$ at some
point. Now consider the integration over $\h$ at this point. We can think of it
as an integral over a holomorphic function $f(\g\h)$, defined on the complex
manifold $\grp SU(2,\CC)$, along a real ``line'' $\{\g\h,\,\h\in\grp SU(2)\}$.
We may shift this real line to $\{\g'\h,\,\h\in\grp SU(2)\}$ without changing
the value of the integral, because the integrand is holomorphic. As a result,
we find that indeed all wave functionals $\Phi_\g$ are equal.

A crucial question is now, whether this is really a state functional or just
the trivial solution $\Psi=0$. Up to now there is no reason why the
integral~\eref{solution} should not vanish. However, it is easy to see that
there are fields $\E mA$ for which $\Phi[\E mA]\ne0$: choose the vierbein such
that $\ee am$ is real, i.e.\ $\E m0=0$. Then the exponent in~\eref{Psi-g-def}
is real too, because $\g^{-1}\del_m\g\in\alg su(2)$ is antihermitian, and
$\Phi$ is given as an integral over a positive real function.

So after all we found exactly one solution to all constraints. However, if we
allow the space manifold to have a non-trivial topology, then there are more
solutions. If there are non-contractible loops, the field $\g$ introduced
in~\eref{A-curv-free} need not be defined globally, and two arbitrary fields
$\g$ and $\g'$ can no longer be transformed into each other by the method just
described. In this case we recover the typical structure of the state space of
three dimensional gravity, where the states can be characterized by the so
called moduli of $\g$, which are in principle the values by which $\g$ is
multiplied after going once around a non-contractible loop. In fact, the
discovery of the formal solution was inspired by a result obtained in three
dimensional gravity, where~\eref{solution} is the general solution to all
constraints, but a crucial difference is that there it can be given as a well
defined object in a different
representation~\cite{matschull:94a,matschull:doc,dewit.matschull.nicolai:93}.

If space time is non-compact, then $\g$ as well as $\h$ must obey certain
boundary conditions. This also leads to topological degrees of freedom for the
field $\g$, namely some kind of soliton numbers, which cannot be gauged away by
transformations of the form~\eref{L-endl}. That the different state functionals
$\Phi_\g$ are parametrized by topological parameters is a first evidence that
$\Phi_\g$ is a vacuum state, as it is typical feature of the vacuum of a
quantum field theory to carry topological degrees of freedom. For different
values of the moduli or different soliton numbers we get different vacua. But
there are still other properties confirming this interpretation.

Perhaps they are even more speculative than the discussion above, but they may
be interesting from a physical point of view. How should a vacuum of quantum
gravity look like? It can certainly not be simply flat space time, as this
would violate the uncertainty relation: all fields would take definit values.
In quantum field theory, a vacuum is usually given by a state that is
annihilated by a set of ``annihilation operators''. This set is, sloppy
speaking, half of the set of all operators and the other half is obtained by
complex conjugation. The question is whether one can recover a similar
structure here. Looking at~\eref{Q-def}, which gives the relation between our
$\P Am$ and the ``real'' momentum $\Q Am$ of the vierbein, one finds that $\P
Am$ looks like, e.g., the annihilation operator of electrodynamics, which is in
principle $E^m+\i B^m=E^m+\i\eps^{mnp}\del_nA_p$, Fourier transformed into
momentum space, where $E$ the electric, $B$ the magnetic field and $A$ the
vector potential. Replacing the vector

 potential by the vierbein and the conjugate momenta, the electric field, by
the real momentum $\Q Am$ shows that $\P Am$ is the analog of the annihilation
operator of electrodymics.

But from this one could infer that the vacuum state is defined by $\P
Am\Psi=0$, i.e.\ $\Psi=1$, which is obviously not a solution to the
constraints. However, in contrast to electrodynamics, $\P Am$ is not covariant,
i.e.\ $\P Am=0$ (as a classical equation) is not invariant under gauge
transformations: remember that $\P Am$ transforms inhomogeneously under
self-dual Lorentz transformations. Thus requiring this to annihilate a state
functional is in contradiction with the constraint equations. To get as close
as possible to the usual definition of a vacuum, one has to look for the
simplest possible holomorphic covariant object that can by build from $\P Am$,
and this is the field strength $\F_{mna}$. Requiring $\F_{mna}\Psi=0$ is
obviously consistent with the constraints because the field strength transforms
covariantly under all local symmetries.

When introducing the field strength $\F_{mna}$ in the beginning as a classical
field, we saw that it is in principle the Riemann curvature tensor in the
self-dual representation. It should be possible to write every ``local''
observable as a function of this tensor, like in electrodymanics where every
observable is a function of the field strengths. Thus we can split the
observables into holomorphic functions of $\F_{mna}$ and holomorphic functions
of $\F\cc_{mna}$, which leads to the typical split into creation and
annihillation operators and the vacuum has the property that it is annihilated
by exactly half of the observables.

In addition, whatever the observables are, if expressed in terms of the Riemann
tensor and thus $\F$, there exists a ``normal ordering'' for the corresponding
quantum operators. All factors of $\F$ are ordered to the right, $\F^*$ to the
left, and any extra vierbein factors in between. As a result, the vacuum
expectation value of any real local observable vanishes. And if there are any
``global'' observables, which cannot be expressed in terms of $\F_{mna}$ but
are functions like parallel transport operators along non-contractible loops
etc., the vacuum expectation values of them will depend on the special vacuum
state, giving the typical structure of a quantum field theory with multiple
vacua.

\section{Other representations and anomalies\label{transit}}
As already mentioned we are somewhere between the metric or Wheeler~DeWitt and
Ashtekar's representation of canonical gravity. The difference to
Wheeler~DeWitt is that we have replaced the metric by a vierbein and added a
imaginary total derivative to the action, leading to complex momenta obeying
reality conditions. In addition, we did not introduce a lapse function nor a
shift vector, but used the ``lower $\t$'' components of the vierbein as
Lagrange multiplier. This gave us a Lorentz covariant expression for the
hamiltonian constraints, which normally splits into a scalar and a vector. The
gauge fixing, however, is identical to that of the usual metric approach: we
only required the spacial hypersurface to be spacelike.

The difference to Ashtekar's representation is that there an additional gauge
fixing of the vierbein must be introduced to make $\A_{pa}$ a ``good''
canonical variable, with a conjugate variable $\ee ap$, the densitized inverse
dreibein. We will see that both can be introduced as phase space functions
here, but without the gauge fixing they are complex and represent more than one
(but less than two) degrees of freedom; i.e.\ a relation like~\eref{reality}
cannot be given for them unless the vierbein takes the usual triangular form.
It is also a consistency check for Ashtekar's formulation of canonical gravity,
that $\A_{pa}$ and $\ee ap$ can be defined such that they obey the basic
Poisson bracket relation without the gauge fixing. Otherwise one could argue
that the Ashtekar's gravity is not fully Lorentz-invariant, because the
variables can only be defined in a gauge fixed version. There are also problems
concerning the diffeomorphism invariance of Ashtekar's
gravity~\cite{matschull:94a}, but we will not gi

ve up the non-singularity of the metric and thus stay on the save side here.

\subsection*{Ashtekar's polynomial representation}
The transition to Ashtekar's representation is rather simple. We already found
that the two quantum operators $\A_{pa}$ and $\ee ap$ are conjugate to each
other when acting on solutions of $\L\cc_a$. There might be extra terms in
their commutators proportional to $\L\cc_a$, if we simply define them as phase
space functions by~\eref{A-von-P-2} and~\eref{ee-def}. To see that they can be
defined such that
\beq[asht-pois]
\pois{\ee am}{\ee bn} = 0, \qquad
\pois{\A_{ma}}{\A_{nb}} =0, \qquad
\pois{\ee am}{\A_{nb}} = \i \eta_{ab} \delta_n^m,
\eeq
we precede as follows. We can write~\eref{P-def}, which is the implicit
definition of $\A_{pa}$ on the constraint surface $\L\cc_a=0$, as
\beq
   \P Am = - \i \, \deldel \ee ap / \E mA / \, A_{pa},
\eeq
where $\ee ap$ is given by~\eref{ee-def}. Now think of $\ee ap$ as some set of
coordinates on the space of all complex rank three matrices $\E mA$. As the
dimension of this space is 12 and $\ee ap$ has only 9 components, we have to
add 3 coordinates, which we will denote by $v\cc_a$. The reason for this
notation is that $\ee ap$ fixes $\E Am$ up to an antiself-dual Lorentz rotation
and $v\cc_a$ is the parameter of that rotation. Now we set
\beq[AP-del]
    \AP pamA = \i \, \deldel \E mA/ \ee ap / , \qquad
     \A_{pa}[E,P] = \i \, \deldel \E mA / \ee ap / \, \P Am .
\eeq
One immediately checks that~\eref{AP-def} is fulfilled, and it is now
straightforward to verify the Poisson brackets above, simply by using the chain
rule for partial derivatives. The phase space function $\A_{pa}$ is still not
unique, as it depends on how the coordinates $v\cc_a$ are chosen.

To get the constraints in the well know form, we define the diffeomorphism and
densitized hamiltonian constraint as
\beq[asht-con]
  \H_m   &=& \E mA \H_A = - \i \ee an \F_{mna}, \zl
  \sqrt{g}\H   &=&  \ft16 \eps^{mnp} \eps_{ABCD}  \E mA \E nB \E pC \H^D =
           \ft12 \eps_{abc} \ee am \ee bn \F_{mnc}.
\eeq
They are formally equal to those of Ashtekar, but remember that $\ee ap$ and
$\A_{pa}$ are not good coordinates on our phase space, as they are complex and
do not obey ``enough'' reality conditions without a Lorentz gauge fixing.
Therefore the Lorentz constraints cannot be expressed in terms of a holomorphic
function of $\ee ap$ and $\A_{ma}$. However, half of them can, and the usual
``Gau\ss\ law'' constraint turns out to be (see~\eref{APrest} for the
definition of $\APrest cBm$)
\beq[asht-lor]
  \L_a - \J_a\^{AB} E_{mA} \APrest cBm \L\cc_c =
   \i \D_m \ee am.
\eeq
The complete recovery of Ashtekar's representation could now be obtained by
gauge fixing ($\E m0=0$) and solving $\L\cc_a$ (for $\P 0m$). But even without
doing this we can use an important result from Ashtekar's representation to
show that our quantum constraint algebra formally closes.

Up to now we did not compute the commutator of $\H_A$ with itself. As we are
not interested in an explicit expression but only want to show that it is again
proportional to the constraints, we invert~\eref{asht-con} and get
\beq
   \H_A = - N_A \sqrt{g} \H  - \ft12 \eps_{ABCD} \eps^{mnp} \E mB \E nC  N^D
\H_p.
\eeq
Now it is obviously sufficient to show that $\sqrt{g}\H$ and $\H_m$ form a
closed algebra, and to do this one needs the brackets~\eref{asht-pois} only.
Thus the calculation is formally equivalent to that in Ashtekar's
representation and we can use the result
from~\cite{brugmann.gambini.pullin:92b}, where it is shown that the quantum
algebra with the operator ordering as in~\eref{asht-con} closes. However, as
this is only a formal result, there might still be anomalies in the algebra
which cannot be discovered by formal calculation. In fact, we will see below
that there is a strong hint for an anomaly, because it is not possible to write
the hamiltonian and diffeomorphism constraints manifestly Lorentz invariant on
a formal level. In other words, it is not possible to write the Lorentz
invariants $\H_m$ and $\H$ as a function of the spacial metric $g_{mn}$ and its
conjugate momentum.

\subsection*{The metric representation}
We will now transform our representation back to the metric or Wheeler~DeWitt
representation, where the wave functional depends on the spacial metric
$g_{mn}$. It is most convenient to do this on the quantum level. We will see
that our vacuum functional can be transformed as well and what we get is a
functional that depends on the spacial metric $g_{mn}$ only.

The first step to recover the metric representation is to choose another
operator for the momentum $\P Am$. Remember that we may define
\beq[new-op]
  \P Am = \i  \deltadelta /\E mA/ - \i \deltadelta \Gf / \E mA / ,
\eeq
with any functional $\Gf[E]$. This operator still obeys the required
commutation relations. The crucial question is: can we find $\Gf$ such that the
Lorentz constraint takes the form
\beq[LAB-metric]
   \L_{AB} = - \i E_{m[A} \deltadelta / \E {B]}m /,
\eeq
so that the inhomogeneous term in~\eref{LAB-def} cancels and $\L_{AB}$
generates Lorentz transformations on the wave functional. Then any solution to
this constraint would be Lorentz invariant and could be expressed in terms of
the spacial metric. To define such a $\Gf$, we have to introduce a dreibein. We
already used the densitized inverse dreibein
\beq
  \ee ap = - \eps^{mnp} \J_{aAB} \E mA \E nB .
\eeq
As we know that this is invertable, we can use it to construct a dreibein
$e_{ma}$, implicitly defined by
\beq[e-def]
  \ee ap = \ft12 \eps^{mnp} \eps_{abc} e_{nb} e_{pc}, \qquad
   e = \det(e_{ma}) = \sqrt{g} >0 .
\eeq
{}From this we can build all other quantities like the inverse dreibein $\e am$
etc. One can also obtain this dreibein by an antiself-dual Lorentz rotation
directly from $\E mA$. If $\Lambda_{AB}$ describes a finite antiself-dual
rotation, then we can always choose it such that $\Lambda_{0A}\E mA=0$, which
fixes $\Lambda_{AB}$ up to a sign. The dreibein is then given by
$e_{ma}=\Lambda_{aA}\E mA$ and fulfills~\eref{e-def}, if we choose the correct
sign for $\Lambda_{AB}$.

We can use the dreibein to define a three dimensional spin connection $\o_{ma}$
via the dreibein postulate
\beq[dreib-post]
  \Dd_{[m} e_{n]a} = \del_{[m} e_{n]a} + \eps_{abc} \o_{[mb} e_{n]c} = 0.
\eeq
This defines a new covariant derivative acting on self-dual indices only.
It can also be given as a functional
derivative~\cite{henneaux.schomblond.nelson:89}
\beq
 \o_{ma} = \deltadelta \Gf / \ee am / , \qquad
  \Gf = \ft12 \intd3\xx \eps^{mnp} e_{ma} \del_n e_{pa} .
\eeq
With this functional inserted above the momentum operator is
\beq
  \P Am =  2\i\eps^{mnp} \J_{aAB} \E nB \o_{pa}+\i  \deltadelta /\E mA/ ,
\eeq
and a short calculation yields~\eref{LAB-metric}. The operator for $\A_{pa}$
becomes
\beq
  \A_{pa} = \o_{pa} -  \deldel \E mA / \ee ap / \deltadelta / \E mA / ,
\eeq
where $\del E /\del\tilde e$ is to be understood as in~\eref{AP-del}.
We can now transform our formal vacuum functional such that it becomes a
solution to the constraints for the new operator representation. For
simplicity, let us drop the topological degrees of freedom here, then we only
have one vacuum state $\Phi=\Phi_{\g=\eins}$. If we define
\beq
  \tildePhi = \exp (\Gf) \, \Phi ,
\eeq
the new operators acting on $\tildePhi$ give the same result as the old
operators acting on $\Phi$, and $\tildePhi$ becomes a formal solution to the
new constraints. By introducing curved Pauli matrices $\sig_m=e_{ma}\,\sig_a$
and using the formula
\beq
   \eps^{mnp}\, \sig_n \sig_p = -2\i \ee am \, \sig_a,
\eeq
we can write $\tildePhi$ in an elegant way as
\beq
  \tildePhi = \int [\d\g] \exp \Big(
   \ft14 \intd3\xx \eps^{mnp}  \Tr\big( (\g \sig_m \g^{-1})
          \del_m (\g\sig_p\g^{-1}) \big) \Big)
\eeq
It is now obvious, at least on a formal level, that this is invariant under
Lorentz rotations, as $\sig_m$ does not transform under antiself-dual rotations
and a self-dual rotation is given by $\sig_m\mapsto\h^{-1}\sig_m\h$, which can
be compensated by a shift in the integration variable $\g\mapsto\g\h$.

Remember that the integral runs over $\grp SU(2)$, but we may shift this
``path'' anywhere in $\grp SU(2,\CC)$. Given a dreibein $e_{ma}$ such that
$g_{mn}$ is real and positive, then we can choose this path such that
$\g\sig_m\g^{-1}$ becomes hermitian for all $\g$, because there allways exists
a rotation that transforms the dreibein into a real dreibein. But then the
integral runs exactly (twice) over all real dreibeins that may be obtained by
rotations from $e_{ma}$. As a result, we can write the vacuum wave functional
in the metric representation as
\beq
  \tildePhi[g_{mn}] = \int [\d e_{ma}] \ \exp \big(\Gf[e_{ma}]\big) ,
\eeq
where the integral runs over all real dreibein fields satisfying
$e_{ma}e_{na}=g_{mn}$, and the measure is assumed to be invariant under $\grp
SO(3)$ rotations.

\subsection*{The Wheeler~DeWitt equation}
An interesting question arising now is: did we find a formal solution to the
``old'' Wheeler~DeWitt equation, as we now have a wave functional
$\tildePhi[g_{mn}]$? To check this we have to write the constraints $\H_m$ and
$\H$ as functional differential equations acting on $\tildePhi[g_{mn}]$. Then
$\H_m$ should generate spacial diffeomorphisms and $\H\tildePhi=0$ should give
the Wheeler~DeWitt equation (see e.g.~\cite{christensen:84}; different relative
factors between the two terms are due to different normalizations of the
action)
\beq[wdw]
     \ft12 \sqrt{g} R[g] \tildePhi - 2  G_{mnpq}
            \frac{\delta^2 \tildePhi }{\delta g_{mn} \delta g_{pq}} =0 ,
\eeq
where R[g] is the spacial curvature scalar and
\beq[G-def]
G_{mnpq}=\ft12 \sqrt{g}^{-1}
           \big( g_{mp} g_{nq} + g_{mq} g_{np} - g_{mn} g_{pq} \big)
\eeq
is the (inverse) ``supermetric'' on the space of three metrics $g_{mn}$.

Let us now assume that the wave functional is given by an arbitrary function
$\tildePsi[g_{mn}]$ of the spacial metric. This is the general solution to
$\L_{AB}\tildePsi=0$. The computation of $\H_m\tildePsi$ and $\H\tildePsi$ is
given in the appendix. The result is rather peculiar and reads
\beq[metr-con]
  \H_m \tildePsi &=&
    2 \, \nabl_{(p} \Big( g_{q)m} \deltadelta \tildePsi/g_{pq}/ \Big). \zl
  \H \tildePsi &=&{} - 2  \sqrt{G}^{-1} \deltadelta /g_{mn}/ \Big(
        \sqrt{G} G_{mnpq} \deltadelta \tildePsi / g_{pq}/ \Big) - \zl
         & &{}-  \Big( \ft12 \sqrt{g} R[g]
          + \ft34\i  \delta(0) \eps^{mnp} e_{ma} \del_n e_{pa}
                 \Big) \,  \tildePsi.
\eeq
The diffeomorphism constraint is exactly what we expected and it requires
$\tildePsi$ to be invariant under spacial diffeomorphisms. However,
$\H\tildePsi$ is different from~\eref{wdw}. First of all, the kinetic term
takes a very nice form: instead of the simple second derivative the Laplace
operator with respect to the ``supermetric'' (whose determinant is $G$)
appears. This is rather surprising, because it came out automatically as a
result of the operator ordering in $\H_A$. In a certain sence our
representation is ``more geometrical'' than the metric representation with the
operator ordering as in~\eref{wdw}.

However, there is an additional divergent term, which is of order $\hbar$ as it
came from a reordering of operators at the same space point. Obviously, this
extra term is not Lorentz invariant; therefore the constraint algebra no longer
closes, and a solution $\tildePsi[g_{mn}]$ can no longer exist. But on the
other hand, we {\em have} the formal solution $\tildePhi$. We must conclude
that something was wrong in our formal calculation. This is a strong hint for
an anomaly in the quantum algebra~\eref{q-const}, which is ``hidden'' in the
ill-defined operator product appearing in $\H_A$. And the fact that $\tildePhi$
formally solves the constraints~\eref{metr-con} and {\em at the same time} is a
functional of $g_{mn}$ results from our assumptions about the measure $[\d\h]$,
which does not exist.

So after all we have to conclude that our construction of the vacuum state was
too formal and maybe such a state functional does not exist because of an
anomaly in the quantum constraint algebra. However, if it is not possible to
define the constraint algebra properly without an anomaly, there is no solution
to the quantum constraints at all, and Dirac's quantization method won't work.
There is a well known example of another diffeomorphism invariant field theory
where exactly this happens, namely string theory. There we know how to quantize
it: the constraints, expressed as the Virasoro generators, split into two
complex conjugate subsets, similar to the split of the observables considered
in \sref{quant}, each forming a closed subalgebra. One defines physical ``wave
functionals'' that are annihilated by one half of the constraints. A state is
then given by equivalence classes: two wave functions are equivalent, if their
difference can be written as some linear combination of the other half of the
constrain

ts, acting on some other wave function.

Is it possible to quantize gravity in a similar way? The crucial question is
whether there is a ``natural'' split of the constraints into two conjugate
subsets, each forming a closed subalgebra when properly regularized. There is
no obvious split of the complete algebra. However, for the Lorentz constraints
we already used this split into $\L_a$ and $\L\cc_a$, which came out
automatically when using the $\alg so(3,\CC)$ representation of the Lorentz
algebra. Therefore a suggestion for a slightly modified Dirac quantization is
to define physical states as follows: Assume that there is a regularization of
the hamiltonian constraints $\H^\epsilon$ such that $\H^0=\H$ as a classical
phase space function, and let us define a set $\Hphys$ of wave functionals
satisfying
\beq[Hphys-def]
   \Psi \in \Hphys  \equivalent \lim_{\epsilon\rightarrow 0}
       \H^\epsilon[\HP^A] \, \Psi = 0 \quad   \forall \HP^A,
\eeq
where the convergence is simply defined pointwise, i.e.\ for any fixed
configuration $\E mA$ we have $\H\Psi[\E mA]\rightarrow0$.

So far this is a regularized Dirac quantization, but now we solve the Lorentz
constraints like the Virasoro constraints in string theory: We solve half of
them, defining a subset $\Lphys\subset\Hphys$ by
\beq
   \Psi \in \Lphys  \equivalent \L[\LP\cc_a] \, \Psi = 0 \quad
           \forall \LP\cc_a
\eeq
and the states $\ket\Psi\in\Sphys=\Lphys/$\relax$\sim$ are defined as
equivalence classes
\beq[Sphys-def]
  \ket {\Psi_1} = \ket {\Psi_2} \equivalent
   \exists \Phi\in\Hphys, \LP_a , \quad
     \Psi_1-\Psi_2 = \L[\LP_a] \Phi.
\eeq
It is essential for this procedure to solve the hamiltonian constraints first,
because otherwise it would be necessary to define $\H^\epsilon$ on equivalence
classes and this requires $\H^\epsilon$ to commute weakly with $\L\cc_a$ {\em
and} $\L_a$. However, if one solves $\H$ first, then for the second step to be
consistent one only needs that the regularized hamiltonian commutes weakly with
$\L\cc_a$. There is a huge class of regularizations with this property, because
only the vierbein in $\H_A$ does not commute with $\L\cc_a$, so for example
every point split regularization like
\beq
  \A_{pa}(\xx) \mapsto \intd3\yy \Delt^\epsilon(\xx,\yy) \A_{pa}(\yy)
\eeq
is of this kind, and $\Delt$ may even be constructed from the spacial metric
$g_{mn}$ providing a diffeomorphism invariant regularization.

In some sense this is just the ``reverse'' of the Ashtekar programme, where one
seeks for a Lorentz invariantly regularized hamiltonian constraint by writing
the field strength as parallel transport matrix along some small loop. There
one is forced to do this, because the Lorentz constraint is solved first. In
addition, the nice complex structure of the Lorentz group does not even appear,
because one is dealing with a gauge fixed representation.

We should emphasize that the Ashtekar programme, which relies on Dirac's
quantization, might fail because of anomalies in the constraint algebra which
cannot be detected by formal calculations, but for which we found some hints
above. Ashtekar's representation also uses another factor ordering, where the
situation is even worse, because there are anomalies already on the formal
level (the structure constants in $[\H,\H]$ appear to the right). In the
quantization programme proposed above such problems do not occur as long as it
is possible to regularize $\H$ such that the equations~\eref{Hphys-def} are
consistent. In particular, for any regularization we have $\Psi_\g\in\Lphys$,
where $\Psi_\g$ is the functional defined in~\eref{Psi-g-def}. Then our vacuum
state (ommitting topological degrees of freedom again)  becomes a perfectly
well defined object
\beq
  \ket{0} = \ket{\Psi_\g}.
\eeq
The ill-defined integration over the local $\grp SU(2)$ has been replaced by
the equivalence class of wave functionals which can be transformed into each
other by Lorentz rotations: the right hand side is in fact independent of $\g$.
The state still has all the properties described in the end of \sref{quant}, so
now we have a well defined vacuum state.

If there is any anomaly, it now appears in the commutator of the regularized
hamiltonian constraint with $\L_a$, as a result of a regularization that is not
Lorentz covariant. What happens is that the equivalence classes become smaller
as they would be without this anomaly, but they remain well defined: given any
state $\ket\Psi\in\Sphys$, then the Lorentz transformed wave function
$\exp(\L[\LP_a])\Psi$ is not necessarily a solution to~\eref{Hphys-def}. It is
just a matter of coincidence that for the special state $\ket0$ every Lorentz
transformed wave function is again a solution to the hamiltonian constraint,
but in general it need not be. However, the equivalence classes themselves and
therefore the states remain Lorentz invariant objects.

As a conclusion, we might summarize the results as follows: So far most works
on quantum gravity with Ashtekar's new variables exploited the fact that the
classical constraints may be written as polynomials in canonically conjugate
variables. One of these variables is the spacial dreibein, whose ``natural''
range is the set of {\em invertable} $3\times3$ matrices. On this space,
however, there is nothing that makes a polynomial behaving better than, e.g., a
rational function, and, as already shown in~\cite{matschull:94a}, extending the
support of the wave function to singular dreibein fields causes some trouble
with the classical limit of quantum gravity as a diffeomorphism invariant
theory. It is the nice geometric structure and not essentially their polynomial
form that makes the constraints more easy to handle.

Another important feature of Ashtekar's variables has not been considered so
much: the representation of the Lorentz group as a complex Lie group. We saw
that it is this structure that makes it possible to define annihilation and
creation operators, normal ordering etc., and to use quantization methods known
from other field theories, which are able to deal with anomalies.
Unfortunately, we were not able to give these operators explicitly, because we
do not know any explicit expression for an observable in quantum gravity; but
what we could do was to give the criteria to classify the observables and a
formal normal ordering prescription. If any such well defined algebra of
observables could be constructed, this would also solve the scalar product
problem: The states are obtained by acting with the creation operators on the
vacuum, providing a Fock space structure. The scalar product can simply be
obtained by defining
$\braket00=1$ and using the commutator algebra of the observables. A definition
of the scalar product as a functional integral over wave functionals is not
required.

\section{Supergravity\label{super}}
In this last and rather technical section we will see that the construction of
the vacuum state can also be made for supergravity. We will give the N=1
example here, but in principle it should be possible to reproduce the same
result for other supergravity models.

\subsection*{Action and constraints}
We introduce a two component Gra\ss mann valued spinor field $\grav_M$, the
gravitino, as superpartner of the vierbein. Definitions for spinors, Pauli
matrices, covariant derivatives etc.\ are given in the appendix, where it is
also shown that the Rarita Schinger action of N=1 supergravity
is~\cite{nieuwenhuizen:81,jacobson:88,matschull:94a}
\beq[super-action]
\Wir[E,\grav] = \ft\i2 \intd4\X \eps^{MNPQ} \, \big(
 4 \, \bgrav_M \sig_N \D_P \grav_Q
   - \E MA \E NB  \J_{aAB}  F_{PQa} \big),
\eeq
and $\O_{MAB}$ is given as a function of the vierbein and the gravitino by the
torsion equation, which again is its own equation of motion
in~\eref{super-action}:
\beq[torsion]
     \D_{[M} \E {N]}A - \bgrav_{[M} \sig^A \grav_{N]} = 0 .
\eeq
The supersymmetry transfomation are parametrized by a spinor field $\seps$ and
read
\beq[susy]
 &\delta \grav_M = -\D_M \seps, \qquad
  \delta \bgrav_M = -\D_M \bseps , &\zl
 & \delta \E MA = \bgrav_M \sig^A \seps - \bseps \sig^A \grav_M .&
\eeq
To show that $\Wir$ is invariant under these transformations one has to use the
torsion equation and the Fierz identities~\eref{fierz}, and one can use the 1.5
order trick, i.e.\ one only has to vary the fields $\E MA$ and $\grav_M$
appearing explicitly in~\eref{super-action}.

Computing the commutator of two supersymmetry transformations, one finds for
the vierbein
\beq
  \comm{\delta_1}{\delta_2} \E MA =
       -  \D_M \big( \bseps_1 \sig^A \seps_2 - \bseps_2 \sig^A \seps_1 \big),
\eeq
which is a translation with $\HP^A=\bseps_1 \sig^A \seps_2 - \bseps_2 \sig^A
\seps_1$. To get the commutator acting on $\grav_M$ we need to know how the
spin connection transforms under supersymmetry. As this is a rather cumbersome
calculation we will not give it here. An easier calculation shows that the
action is invariant under the following translations:
\beq[super-trans]
  \delta \E MA = -\D_M \HP^A , \qquad
    \delta \grav_M &=& 2 \HP^N \, \D_{[M} \grav_{N]} , \zl
    \delta \bgrav_M &=& 2 \HP^N \, \D_{[M} \bgrav_{N]}.
\eeq
We see that in supergravity the translations rather than the diffeomorphisms
generated by the Lie derivatives appear as the basic symmetries: they are the
commutators of two local supersymmetry transformations, and thus the local
versions of the Poincar\'e translations.

Let us now set up the canonical formulation. The space time split leads to the
following lagrangian, whose ``bosonic'' part is formally equal to~\eref{lag-2}:
\beq[lag-bos]
    \Lag_{\rm bos} = \intd3\xx  \i\eps^{mnp} \Big(
 2 \, \del_\t\E mA\,\E nB\J_{aAB}\A_{pa} - \ \ \ \ \ \ && \zl
  -  \D_p \big( \E mA \E nB \J_{aAB} \big) \,\A_{\t a}
 &-&   \E \t A \E mB \J_{aAB} \F_{npa}   \Big),
\eeq
and the part containing the gravitino explicitly is
\beq[lag-super]
    \Lag_{\rm ferm} =  \intd3\xx 2\i\eps^{mnp}\,\Big(
                  \bgrav_m \sig_n \del_\t \grav_p
   - \ft\i2 \A_{\t a} \, \bgrav_m \sig_n \sig_a \grav_p  &-&\zl
  -  \bgrav_m \sig_\t \D_n \grav_p \
     +  \bgrav_\t \sig_m \D_n \grav_p
         &+&  \D_p \big( \bgrav_m \sig_n \big) \grav_\t  \Big).
\eeq
{}From the velocity terms we obtain the momenta
\beq[P-super-def]
\P Am  = 2 \i \eps^{mnp}  \J_{aAB} \E nB \A_{pa}, \qquad
\pgrav^m = - 2\i \eps^{mnp} \, \bgrav_n \sig_p .
\eeq
The canonical variables are $\E mA$, $\P Am$, $\grav_m$ and $\pgrav^m$. The
equation above for $\pgrav^m$ can be inverted to give $\bgrav_m$ as a phase
space function, provided that the spacial metric is invertable, and $\A_{pa}$
is again given by~\eref{A-von-P-2}. The basic Poisson brackets are
\beq
\pois{ \E mA }{ \P Bn } = \delta_m^n \, \delta_B^A,
\qquad \pois{ \grav_m } { \pgrav^n } = -   \delta_m^n \, \eins.
\eeq
Note that when differentiating $\Lag$ with respect to $\del_\t \grav_m$ we get
an additional minus sign by anticommuting the derivative operator with
$\bgrav_m$, and that the Poisson brackets are symmetric for Gra\ss mann valued
entries.

The reality conditions for $\P Am$ are slightly more complicated now, as there
is a contribution from the gravitinos in the torsion equation:
\beq[super-reality]
\Pcc Am = \P Am - 2 \i \eps^{mnp} \, \big(\del_n E_{pA}
                                 - \bgrav_n \sig_A \grav_p \big).
\eeq
The Lorentz constraint $\L\cc_a$, however, remains unchanged, as it follows
from the unchanged equation for $\P Am$ as a function of $\A_{pa}$. But the
complex conjugate $\L_a$ is different, since we have a new
relation~\eref{super-reality}. The rest of the constraints is obtained by
differentiating $\Lag$ with respect to $\E \t A$, $\grav_\t$ and $\bgrav_\t$,
and of course they are the generators of local supersymmetry and translations.
The complete set of constraints is
\beq[super-con]
\L\cc_a   &=& \J\cc_a\^A\_B \, \P Am  \E mB, \zl
\L_a      &=& \J_a\^A\_B \, \P Am  \E mB - \i  \del_m \ee am
             -\ft\i2  \, \pgrav^m \sig_a \grav_m , \zl
\S        &=& 2 \i \eps^{mnp} \, \sig_m \D_n \grav_p , \zl
\bS       &=& - \D_m \pgrav^m, \zl
\H_A      &=& -\i \eps^{mnp}  \J_{aAB}  \E mB  \F_{npa}
           - 2\i \eps^{mnp} \, \bgrav_m \sig_A \D_n \grav_p .
\eeq
Note that, as $\grav_m$ and $\pgrav^m$ both transform under the self-dual
representation of the Lorentz algebra, all these constraints are holomorphic in
$\A_{pa}$. The smeared versions of the bosonic constraints are defined as
usual. For the fermionic ones we set
\beq
\S[\bseps] = \intd3\xx \bseps\S, \qquad
\S[\seps] = \intd3\xx \bS \seps.
\eeq

\subsection*{Solving the quantum constraints}
For the quantum theory we will choose the $E$-$\pgrav$-representation. The wave
functional becomes a function $\Psi[\E mA,\pgrav^m]$, and the momentum
operators are
\beq[super-op]
\P Am(\xx) = \i \deltadelta / \E mA(\xx) / , \qquad
\grav_m(\xx)  =  \i \deltadelta / \pgrav^m(\xx) /.
\eeq
Again, $\L\cc_a$ is solved, if $\Psi$ depends on $\E mA$ only via a holomorphic
function of $\ee ap$, and on such a $\Psi[\ee am,\pgrav^m]$ the spin connection
acts as
\beq
 \A_{pa} \, \Psi = -\deltadelta \Psi / \ee ap /  .
\eeq
We can now solve $\H_A$ and $\S$ in the same way as we solved $\H_A$ for the
bosonic theory. We look for simple solutions which are already annihilated by
the field strength $\F_{mna}$ and the ``super field strength''
$\D_{[m}\grav_{n]}$. Again, the general solution is well known and similar to
the general solution of three dimensional
supergravity~\cite{dewit.matschull.nicolai:93}. For vanishing fields strengths
the gauge fields are locally given by an $\grp SU(2,\CC)$ field $\g$ and a
spinor field $\fmod$ and read
\beq
   \A_{ma} = \i \Tr(\g^{-1} \del_m \g \sig_a ) , \qquad
   \grav_m = \g^{-1} \del_m \fmod.
\eeq
The corresponding quantum eigenvalue equations can easily be solved, and the
result is
\beq
   \Psi_{\g,\fmod} = \exp \Big\{ - \i \intd3\xx \big(
         \Tr(\g^{-1} \del_m \g \sig_a)\,  \ee am
       + \pgrav^m \g^{-1} \del_m \fmod \big) \Big\} .
\eeq
Finally we have to solve $\L_a$ and $\bS$. As in~\eref{L-endl} we exponentiate
these constraints and get~\cite{matschull:doc,dewit.matschull.nicolai:93}
\beq
\exp\big(\L[\LP_a]\big) \, \Psi_{\g,\fmod} &=& \Psi_{\g\h,\fmod}, \quad
     {\rm where} \quad
\h = \exp\big( -\ft\i2  \LP_a \, \sig_a\big), \zl
\exp\big(\S[\seps]\big) \, \Psi_{\g,\fmod} &=& \Psi_{\g,\fmod+\g\seps}.
\eeq
To get formal solution to all constraints, we have to integrate this
expressions over $\seps$ and $\h$:
\beq[super-solution]
\Phi_{\g,\fmod} = \int [\d\h] \int [\d\seps] \, \Psi_{\g\h,\fmod+\g\seps},
\eeq
where we have to assume that the measure is a Haar measure on $\grp SU(2)$ and
the spinor measure is invariant under $\grp SU(2)$ rotations of $\seps$. These
solutions have the same properties as those given in~\eref{solution} for the
bosonic theory: they carry topological degrees of freedom, are annihilated by
half of the observables etc., and they may be regarded as the vacuum states of
supergravity for the same reasons.

Of course, regarding ill-defined operator products and functional integrals the
same as for the bosonic theory holds for supergravity. If there is an anomaly
in the constraint algebra, we have to choose another quantization method.
However, we can precede in exactly the same way as in \sref{transit}, as not
only the Lorentz constraints but also the supersymmetry generators split into
two conjugate subsets, so instead of integrating over $\h$ and $\fmod$, we
define states as equivalence classes, just replacing $\L[\LP_a]\Phi$ by
$\L[\LP_a]\Phi+\S[\seps]\Phi^\prime$ in~\eref{Sphys-def}. Again we obtain a
well defined vacuum state
\beq
  \ket 0 = \ket{\Psi_{\g,\fmod}}.
\eeq

\appendix
\section*{Appendix}
\section{$\J$-symbols, Lorentz algebra, and Pauli matrices}

The $\J$-symbols were introduced in~\eref{alg-map} as the algebra isomorphism
between the two representations of the Lorentz algebra $\alg so(3,\CC)$ and
$\alg so(1,3)$. Here we summerize the properties of these symbols and give some
formulas (see~\cite{matschull:94a,matschull:doc} for more details on this
notation). They provide a complete and orthonormal (complex) basis of $\alg
so(1,3)$, i.e.\ they are antisymmetric in $\scr A,\scr B$ and
\beq[J-complete]
&\J_a\^{AB} \J_{bAB} = \eta_{ab}, \qquad
\J\cc_a\^{AB} \J\cc_{bAB} = \eta_{ab}, \qquad
\J_a\^{AB} \J\cc_{bAB} = 0 , &\zl
&\J_a\^{AB} \J_{aCD} + \J\cc_a\^{AB} \J\cc_{aCD} = \delta^{AB}_{CD},&
\eeq
where $\delta^{AB}_{CD}=\ft12\delta^A_C\delta^B_D-\ft12\delta^A_D\delta^B_C$
and $\eta_{ab}=\delta_{ab}$ is the metric on $\alg so(3)$. Note that this is
just the ``spatial'' part of $\eta_{AB}$, where the indices take the values
$1,2,3$ only, and using the same symbol will be useful below.

Furthermore, the $\J$s have to respect the Lie algebra structure. In fact, they
also provide two four dimensional Clifford representation of $\alg so(3)$,
\beq[so3-rep]
\J_{aA}\^B \J_{bB}\^C   &=& -\ft14 \eta_{ab}\delta_A\^C
                        +\ft12 \eps_{abc}\J_{cA}\^C, \zl
\J\cc_{aA}\^B \J\cc_{bB}\^C   &=& -\ft14 \eta_{ab}\delta_A\^C
                        +\ft12 \eps_{abc}\J\cc_{cA}\^C,
\eeq
commuting with each other:
\beq[J-comm]
\J_{aA}\^B \J\cc_{bB}\^C = \J\cc_{bA}\^B \J_{aB}\^C.
\eeq
An explicit representation is given by
\beq[J-def]
\J_{aAB} = \ft\i2 \eta_{aA} \delta_B\^0 - \ft\i2 \eta_{aB} \delta_A\^0
           - \ft12 \eps^0\_{aAB},
\eeq
where $\eps^{ABCD}$ is the four dimensional Levi Civita symbol defined by
$\eps^{0123}=-\eps_{0123}=1$. Observe that the three dimensional symbol, which
gives the structure constants of $\alg so(3)$ in~\eref{so3-rep}, is obtained by
$\eps_{abc}=\eps^0\_{abc}$. Note also that~\eref{J-def} defines the generator
$\J_a$ as the combination ``rotation around $a$-axis~$+$~$\i\times$boost in
$a$-direction'', which splits the Lorentz algebra into its self-dual and
antiself-dual part. Indeed, we find that
\beq[self-dual]
\eps_{AB}\^{CD}\J_{aCD} = 2\i \J_{aAB}, \qquad
\eps_{AB}\^{CD}\J\cc_{aCD} = - 2\i \J\cc_{aAB}.
\eeq
By dropping the $\J^*$ from the sum in~\eref{J-complete} we obtain
\beq[sd-proj]
  \J_a\^{AB} \J_{aCD} = \ft12 \delta^{AB}_{CD} - \ft\i4 \eps^{AB}\_{CD},
\eeq
which is the projector onto the self-dual part of an antisymmetric tensor.
We can now define the Lorentz-covariant derivatives of various objects carrying
different kinds of ``flat'' indices. For a 4-vector $V^A$ we have
\beq[cov-vec]
\D_M V^A = \del_M V^A + \O_M\^A\_B V^B.
\eeq
An antisymmetric tensor $T^{AB}$ can be transformed into a ``self-dual''
3-vector $T_a=\J_{aAB}T^{AB}$, whose derivative reads
\beq[cov-tens]
\D_M T_a = \del_M T_a + \eps_{abc} \A_{Mb} T_c, \qquad
\D_M T\cc_a = \del_M T\cc_a + \eps_{abc} \A\cc_{Mb} T\cc_c.
\eeq
Note that we are using the same symbols $a,b,\dots$ for both indices
transforming under the self-dual and under the antiself-dual representation of
the Lorentz algebra. As mixed tensors will not appear throughout this article,
a tensor with a $*$ always carries antiself-dual indices. The special tensors
$\eta$, $\eps$, $\J$ and $\J^*$ are constant under the covariant derivative.

The field strength of the spin connection is defined as usual via the
commutator of two covariant derivatives and can be given in both the $\alg
so(1,3)$ representation
\beq[R-def]
\R_{MNAB} = \del_M \O_{NAB} - \del_N \O_{MAB}
            + \O_{MA}\^C \O_{NCB} - \O_{NA}\^C \O_{MCB},
\eeq
or in the $\alg so(3,\CC)$ representation
\beq[F-def]
\F_{MNa} = \del_M \A_{Na} - \del_N \A_{Ma}
            + \eps_{abc} \A_{Mb} \A_{Nc}.
\eeq
Of course, they are related by $\F_{MNa}=\J_a\^{AB}\R_{MNAB}$. If, in addition,
the spin connection is given by the vierbein postulate~\eref{vierb-post}, then
they are related to the Riemann tensor by $R_{MNPQ}=\E PA \E QB \R_{MNAB}$. To
express the Einstein Hlbert action in terms of $\A_{Ma}$, we use the definition
\beq
  E \, \eps^{ABCD} = \eps^{MNPQ} \E MA \E NB \E PC \E QD
\eeq
of the vierbein determinant to obtain
\beq
\Wir[E]   &=& \ft12 \intd4\X E R = \ft12  \intd4\X E \, \E CP \E DQ
\R_{PQ}\^{CD} \zl
  &=&  -\ft18 \intd4\X \eps^{MNPQ} \eps_{AB}\^{CD} \E MA \E NB \R_{PQCD}.
\eeq
To this expression we add the ``square'' of the vierbein postulate. This
vanishes identically {\em and} does not change the equation of motion for
$\O_{MAB}$, i.e. if we add
\beq
 -\ft\i2 \intd4\X   \eps^{MNPQ} \D_M \E NA \, \D_P E_{QA},
\eeq
then the action is still 1.5 order in $\O_{MAB}$. After a partial integrations
one finds that this is equal to
\beq
 -\ft\i4 \intd4\X \eps^{MNPQ}    \E MA \E NB \R_{PQAB}
\eeq
Using~\eref{sd-proj} we find
\beq
\Wir[E]  &=& -\ft\i2 \intd4\X \eps^{MNPQ} \E MA \E NB \J_{aAB} \J_a\^{CD}
\R_{PQCD} \zl
   &=& -\ft\i2 \intd4\X \eps^{MNPQ} \E MA \E NB \J_{aAB} \F_{PQa},
\eeq
which gives the action~\eref{action}.

For supergravity we have to define Majorana fermions. For our purpose it is
most useful to represent them as two component Gra\ss mann valued complex
spinors $\grav$. The conjugate spinor is defined by $\bgrav=\i\grav^\dagger$.
Under local Lorentz rotations they transform under the (anti)self-dual
representation, i.e.
\beq
\D_M \grav = \del_M \grav - \ft\i2 \A_{Ma} \, \sig_a \grav,  \qquad
\D_M \bgrav = \del_M \bgrav + \ft\i2 \A\cc_{Ma} \, \bgrav \sig_a ,
\eeq
where $\sig_a$ are the Pauli matrices. To build a vector bilinear from a spinor
(this is the only bilinear we need), we have to provide four dimensional
``gamma matrices'' such that $\bgrav_1 \sig_A \grav_2$ transforms as a vector.
This can be achieved by using the same Pauli matrices together with
$\sig_0=\eins$. One finds the algebra~\cite{matschull:94a}
\beq
  \sig_A \sig_a = 2\i \J_{aAB} \, \sig^B, \qquad
  \sig_a \sig_A = -2\i \J\cc_{aAB} \, \sig^B.
\eeq
Note that the three dimensional index $a$ is a self-dual index if it appears to
the right of $\sig_A$ but an antiself-dual index if it appears on the left. As
the gamma matrices are hermitian, the vector bilinear obeys
\beq
  \big( \bgrav_1 \sig_A \grav_2 \big)^* =
  - \bgrav_2 \sig_A \grav_1.
\eeq
The supersymmetric partner of the vierbein is the gravitino $\grav_M$, and the
Rarita Schwinger action for supergravity~\cite{nieuwenhuizen:81,jacobson:88}
takes its simplest form in 1.5 order formalism, where the spin connection
$\O_{MAB}$ (and therefore $\A_{Ma}$) is defined by the torsion equation
\beq[super-torsion]
     \D_{[M} \E {N]}A - \bgrav_{[M} \sig^A \grav_{N]} = 0 .
\eeq
It is the equation of motion for $\O_{MAB}$ in
\beq[RS-action]
\Wir[E,\grav] = \intd4\X \eps^{MNPQ} \, \Big(
     &-&\ft18 \eps_{AB}\^{CD} \, \E MA \E NB \, \R_{PQCD} + \zl
 &+&\i\,\bgrav_M \sig_N \D_P \grav_Q - \i\,\D_M \grav_N \sig_P \grav_Q \Big),
\eeq
with $\sig_N=\E MA\, \sig_A$. Here the fields strength $\R_{PQCD}$ is no longer
the Riemann tensor with flat indices, since the vierbein postulate has been
replaced by the torsion equation. One can again write the action in terms of
$\A_{Ma}$ by adding the ``square'' of the torsion equation
\beq
 -\ft\i2 \intd4\X
        \eps^{MNPQ} \big( \D_M \E NA - \bgrav_M \sig^A \grav_N \big) \,
                        \big( \D_P E_{QA} - \bgrav_P \sig_A \grav_Q \big)
\eeq
to the action. This is equal to
\beq
\intd4\X \eps^{MNPQ} \, \Big(
     &-&\ft\i4  \, \E MA \E NB \, \R_{PQAB} + \zl
  &+& \i\,\bgrav_M \sig_N \D_P \grav_Q + \i\,\D_M \grav_N \sig_P \grav_Q \Big),
\eeq
and adding it to~\eref{RS-action} gives~\eref{super-action}. The vanishing of
the quartic term in $\grav$ is a due to the Fierz identities for Gra\ss mann
valued spinors. The formula needed here is
\beq[fierz]
  \bgrav_1 \sig_A \grav_2 \, \bgrav_3 \sig^A \grav_4 =
  \bgrav_1 \sig_A \grav_4 \, \bgrav_3 \sig^A \grav_2,
\eeq
and other useful identities are
\beq
 \J_{aAB}\,  \bgrav_1 \sig^A \grav_2 \, \bgrav_3 \sig^B \grav_4 &=&
 \J_{aAB}\, \bgrav_1 \sig^A \grav_4 \, \bgrav_3 \sig^B \grav_2, \zl
 \J\cc_{aAB}\,  \bgrav_1 \sig^A \grav_2 \, \bgrav_3 \sig^B \grav_4 &=&
 -\J\cc_{aAB}\, \bgrav_1 \sig^A \grav_4 \, \bgrav_3 \sig^B \grav_2.
\eeq

\section{Recovering the Wheeler~DeWitt equation\label{wdw-comp}}
Here the transformation of the hamiltonian constraints $\H_A$ to the metric
representation defined in \sref{transit} will be carried out. We assume that
the wave functional is a function $\tildePsi[g_{mn}]$ of the spacial metric,
which is the general solution to the Lorentz constraints.

We introduced the three dimensional complex spin connection $\o_{ma}$ via the
dreibein postulate~\eref{dreib-post}. To simplify notation, we define an
operator
\beq
    p_{pa} = \i\o_{pa}- \i \A_{pa}.
\eeq
This looks like the well known split of Ashtekar's variables into the spin
connection and the momentum of the dreibein, but remember that $\o_{pa}$ as
well as $\p_{pa}$ are complex here and do not represent real and imaginary part
of $\A_{pa}$. However, for the commutator of $\p_{ma}$ and $\ee am$ we find the
canonical relations (remember that spacial delta function are to be restored)
\beq[p-e-comm]
  \comm{\p_{ma}}{\ee bn} =  \i \eta_{ab} \delta_m^n, \qquad
  \comm{\p_{ma}}{\p_{nb}} =  0,
\eeq
and on a holomorphic functional of the dreibein $\p_{pa}$ acts as
$\i\delta/\delta\ee ap$. As $g_{mn}$ is such a functional, defined
in~\eref{g-von-ee}, we find
\beq[p-auf-g]
  \p_{pa} \tildePsi = \i \, \deltadelta \tildePsi/ \ee ap / =
       \i\, \deldel g_{mn} / \ee ap / \deltadelta \tildePsi/   g_{mn} /.
\eeq
Explicitly we have
\beq[del-g-del-ee]
\deldel g_{mn} / \ee ap / = e^{-1} \big(
     g_{mn} e_{pa} - g_{mp} e_{na} - g_{np} e_{ma} \big)
     = - 2 G_{mnpq} \e aq,
\eeq
where $G_{mnpq}$ is the inverse supermetric introduced in~\eref{G-def}.
The field strength appearing in the constraints becomes
\beq[F-metr]
  \F_{mna}   &=& \R_{mna} + 2 \i \del_{[m} \p_{n]a}
           + \i \eps_{abc} ( \o_{mb} \p_{nc} + \p_{mb} \o_{nc} )
           - \eps_{abc} \p_{mb} \p_{nc} \zl
  &=& \R_{mna} + 2 \i \Dd_{[m} \p_{n]a}  - \eps_{abc} \p_{mb} \p_{nc}
      + \i \eps_{abc}  \comm{p_{mb}}{\o_{nc}},
\eeq
where
\beq
\R_{mna} = \del_m \o_{na} - \del_n \o_{ma} + \eps_{abc} \o_{mb} \o_{nc}
\eeq
is the field strength of $\o_{ma}$. It is related to the three dimensional
Riemann tensor by
\beq
R_{mnpq} = \eps_{abc} \R_{mna} e_{bp} e_{cq}.
\eeq
Note that this is again real because it is the Riemann tensor of a real metric
although the dreibein and the spin connection are complex. Using the symmetries
of the Riemann tensor we find the useful relations
\beq[3-R]
  \eps^{mnp} \R_{mna} e_{pa} = e R, \qquad
  \R_{mna} \ee am = 0,
\eeq
where now $R$ denotes the three dimensional Ricci scalar.

In~\eref{F-metr} we picked up a singular term when ordering the spin connection
in $\Dd_m \p_{na}$ to the left, using the commutator~\eref{p-e-comm}: the two
entries of the commutator are to be taken at the same point. The diffeomorphism
constraint, given in~\eref{asht-con}, now reads (the term containing $\R_{mna}$
vanishes by~\eref{3-R})
\beq[diff-metr]
 \H_m = 2\, \nabl_{[m}  \big( \ee an  \p_{n]a} \big)
         +\i \eps_{abc} \p_{mb} \ee an \p_{nc}
         + \i  \eps_{abc} \ee an
                 \comm{p_{mb}}{\o_{nc}}.
\eeq
To get the first term we used the fact that the dreibein is covariantly
constant under $\Dd_m$: here $\nabl_m$ is the metric covariant derivative with
respect to the spacial metric $g_{mn}$. In the second term we placed the
dreibein factor between the two $p$-factors, which is allowed
by~\eref{p-e-comm}. One can either show that this gives a term proportional to
the Lorentz constraint or use~\eref{p-auf-g} to see that it vanishes when
acting on $\tildePsi[g_{mn}]$.

The singular term also drops out (formally, but that's all we can say without a
proper regularization). By~\eref{p-e-comm} we can put the dreibein inside the
commutator and get
\beq
\i  \eps_{abc}  \comm{p_{mb}}{\o_{nc} \ee an} =
  - \i  \comm{p_{mb}}{ \del_n \ee bn}
\eeq
Splitting the points where the fields are taken, we get an expression like
$\delta(\xx,\yy)\del_m\delta(\xx,\yy)$, integrated over $\yy$, and this
vanishes because it is a total divergence. A similar argument is also used
in~\cite{brugmann.gambini.pullin:92b} to show that $\H_m$ generates
diffeomorphisms in Ashtekar's representation.

The only term surviving in~\eref{diff-metr} is the first one.
With~\eref{del-g-del-ee} and after some algebra we get
\beq
  \H_m \tildePsi = 2 \, \nabl_{(p} \Big( g_{q)m} \deltadelta \tildePsi/g_{pq}/
                   \Big).
\eeq
This just requires $\tildePsi$ to be invariant under spacial diffeomorphisms
$\delta g_{mn}=\nabl_{(m} v_{n)}$ with some vector field $v_n$.

The hamiltonian constraint has also been given in~\eref{asht-con}. Without the
extra density factor it reads
\beq[H-metr]
  \H &=&\ft12 \eps^{mnp} e_{ma} \F_{npa} \zl
     &=&\ft12 e R  + \i \eps^{mnp} \del_n \big( e_{ma} \p_{pa} )
        - \ft12 \eps^{mnp} \eps_{abc} e_{ma} \p_{nb} \p_{pc} + \zl
       &&  {}+\ft\i2 \eps^{mnp} \eps_{abc} e_{ma}
                        \comm{\p_{nb}}{\o_{pc}}.
\eeq
Here we used~\eref{3-R} for the first term and the dreibein
postulate~\eref{dreib-post} to get the second term. The first term is already
the required ``potential term'' in~\eref{wdw}. The second term vanishes when
acting on $\tildePsi$ as
\beq
   \eps^{mnp} e_{ma} p_{pa} \tildePsi =
   \i \eps^{mnp} e_{ma} \deldel g_{rs} / \ee ap /
         \deltadelta \tildePsi/g_{rs}/
   = 0,
\eeq
which follows immediately from~\eref{del-g-del-ee}. The third term is quadratic
in $\p_{ma}$ and yields the ``kinetic part'' of the Wheeler~DeWitt equation. We
rewrite it using~\eref{p-auf-g} as
\beq[kin-term]
  - \ft12 \eps^{mnp} \eps_{abc} e_{ma} \p_{nb} \p_{pc} \tildePsi
  &=&   \ft12  \deldel \ee bn / e_{pc} / \deltadelta / \ee bn /
     \Big( \deldel g_{rs} / \ee cp / \deltadelta \tildePsi / g_{rs} / \Big) \zl
   &=& -  \deltadelta / e_{pc} /
     \Big(G_{rspq} \e cq \, \deltadelta \tildePsi / g_{rs} / \Big) \zl
   &=& - 2  \sqrt{g} \deltadelta / g_{pq} /
     \Big(\sqrt{g}^{-1} G_{mnpq} \, \deltadelta \tildePsi / g_{mn} / \Big).
\eeq
To get the last line we used the following formula which holds for any function
$f(g_{mn})$:
\beq
  \deldel / e_{pc} / \Big( f \e cq \Big)
    = 2 \sqrt{g} \, \deldel / g_{pq} / \Big( f \sqrt{g}^{-1} \Big).
\eeq
To show that what we got is indeed the Laplace operator appearing
in~\eref{metr-con}, we only have to show that the determinant of the
supermetric $G$ is proportional to $g^{-1}$. As $G_{mnpq}$ is the inverse
supermetric (the ``coordinates'' $g_{mn}$ have lower indices), its determinant
should be proportional to $g$. That this is in fact true can be seen as
follows: we know that the determinant of $G_{mnpq}$ is given as a homogeneous
polynomial of degree 6, and from~\eref{G-def} we infer that it is a polynomial
of degree 12 in $g_{mn}$, multiplied by $g^{-3}$. However, there is only one
polynomial of degree 12 in $g_{mn}$ that gives a scalar under spacial
diffeomorphisms, and this is $g^4$, so all together we have
$\det(G_{mnpq})\propto g$ and $G\propto g^{-1}$.

Finally we have to consider the singular term
\beq
 \ft\i2 \eps^{mnp} \eps_{abc} e_{ma} \comm{\p_{nb}}{\o_{pc}}
= \ft\i2 \eps^{mnp} \big( \comm{\p_{nb}}{\del_m e_{pb}}
                        - \eps_{abc} \comm{\p_{nb}}{e_{ma}} \o_{pc} \big).
\eeq
Here we need the commutator
\beq
  \comm{\p_{nb}}{e_{ma}} = \i \deldel e_{ma}/\ee bn/
     = \i e^{-1} \big(\ft12 e_{ma}e_{nb} - e_{mb} e_{na}\big).
\eeq
Contacting $a,b$ gives something symmetric in $m,n$, so the first term above
vanishes. The second gives
\beq
  - \ft34 \i  \delta(0) \eps^{mnp} \eps_{abc} e_{ma} e_{nb} \o_{pc}
 = - \ft34 \i  \delta(0) \eps^{mnp} e_{ma} \del_n e_{pa},
\eeq
where $\delta(0)=\delta(\xx,\xx)$ is the infinite constant we get from
commuting to operators at the same point. Adding all the results together we
get~\eref{metr-con}.


\begin{thebibliography}{10}
\bibitem{ashtekar:86}
Ashtekar A 1986
 New variables for classical and quantum gravity.
 {\em Phys. Rev. Lett.} {\bf 57} 2244
\bibitem{jacobson.smolin:88b}
Jacobson  T and Smolin L 1988
 Covariant action for Ashtekar's form of canonical gravity.
 {\em Class. Quant. Grav.} {\bf 5} 583
\bibitem{matschull.nicolai:92}
Nicolai H and Matschull H-J 1993
 Aspects of canonical gravity and supergravity.
 {\em Jour. Geom. Phys.} {\bf 11} 15
\bibitem{matschull:94a}
Matschull H-J 1994
 On loop states in quantum gravity and supergravity.
 {\em Class. Quant. Grav.} {\bf 11} 2395
\bibitem{jacobson.smolin:88}
Jacobson  T and Smolin L 1988
 Nonpertubative quantum geometries.
 {\em Nucl. Phys.} B {\bf 299} 295
\bibitem{dirac:65}
Dirac P A M 1965
 {\em Lectures on Quantum Mechanics}.
 (New York, Academic)
\bibitem{wheeler:64}
Wheeler J A 1964
 Geometrodynamics and the issue of the final state.
 In C.~DeWitt and B.~DeWitt, editors, {\em Relativity, Groups and
  Topology}. (New York: Gordon and Breach)
\bibitem{dewitt:67}
DeWitt B 1967
 Quantum theory of gravity I,II.
 {\em Phys. Rev.} {\bf 160} 1113; {\bf 162} 1195
\bibitem{fukuyama.kamimura:90}
Fukuyama T and Kamimura K 1990
 Complex action and quantum gravity.
 {\em Phys. Rev.} D {\bf 41} 1105
\bibitem{henneaux.schomblond.nelson:89}
Henneaux M, Schomblond C and Nelson J E 1989
 Derivation of Ashtekar's variables from tetrad gravity.
 {\em Phys. Rev.} D {\bf 39} 434
\bibitem{matschull:doc}
Matschull H-J 1994
 Kanonische Formulierung von Gravitations- und Super\-gravi\-tations-Theorien.
 {\em Dissertation} (Universit\"at Hamburg) DESY 94-118
\bibitem{dewit.matschull.nicolai:93}
de~Wit B, Matschull H-J and Nicolai H 1993
 Physical states in d=3, N=2 supergravity.
 {\em Phys. Lett.} {\bf B318} 115
\bibitem{brugmann.gambini.pullin:92b}
Br\"ugmann B, Gambini R, and Pullin J 1992
 Jones polynomials for intersecting knots as physical states of
  quantum gravity.
 {\em Nucl. Phys.} B {\bf 385} 587
\bibitem{christensen:84}
Christensen S M 1984
 {\em Quantum Theory of Gravity}.
 (Bristol: Adam Hilger)
\bibitem{nieuwenhuizen:81}
van Nieuwenhuizen P 1981
 Supergravity.
 {\em Phys. Rep.} {\bf 68} 189
\bibitem{jacobson:88}
Jacobson T 1988
 New variables for canonical supergravity.
 {\em Class. Quant. Grav.} {\bf 5} 923
\end{thebibliography}
\end{document}